\documentclass[apj,twocolumn]{emulateapj}
\usepackage{epsfig,apjfonts,mathptmx}

\newcommand{\mt}{\hbox{$M_{\rm 20}$}}

\newcommand{\Kv}  {\hbox{$K_{\rm Vega}$}}

\newcommand{\zph}  {\hbox{$z_{\rm phot}$}}

\newcommand{\gsim}{\lower.5ex\hbox{$\; \buildrel > \over \sim \;$}}
\newcommand{\lsim}{\lower.5ex\hbox{$\; \buildrel < \over \sim \;$}}
\newcommand{\SpT}{{\sf SpT}}
\newcommand{\dS}  {\hbox{$d_{\rm S}$}}
\newcommand{\dI}  {\hbox{$d_{\rm I}$}}
\newcommand{\dG}  {\hbox{$d_{\rm G}$}}
\newcommand{\dGA}  {\hbox{$d_{\rm GA}$}}
\newcommand{\dGM}  {\hbox{$d_{\rm GM}$}}

\shorttitle{EXTREMELY RED OBJECTS IN COSMOS}
\shortauthors{Kong et al.}

\begin{document}

\title{Classification of Extremely Red Objects in the COSMOS Field}

\author{
Xu Kong\altaffilmark{1,2},
Guanwen Fang\altaffilmark{2},
Nobuo Arimoto\altaffilmark{3},
Min Wang\altaffilmark{2}
}
\altaffiltext{1}{Key Laboratory for Research in Galaxies and 
Cosmology, University of Science and Technology of China, 
Chinese Academy of Sciences, China}
\email{xkong@ustc.edu.cn}

\altaffiltext{2}{Center for Astrophysics, University of Science and
Technology of China, Hefei, Anhui, 230026, China}

\altaffiltext{3}{Optical and Infrared Astronomy Division, National
Astronomical Observatory, Mitaka, Tokyo 181-8588, Japan}

\begin{abstract}

We present a study of the classification of $z \sim 1$
extremely red objects (EROs), using a combination of 
$Hubble$ $Space$ $Telescope$ ($HST$) Advanced Camera 
for Surveys (ACS), $Spitzer$ Infrared Array Camera
(IRAC), and ground-based images of the COSMOS field. Our sample
includes $\sim 5300$ EROs with $ i-K_\mathrm{s} \ge 2.45$ (AB,
equivalently $I-K_\mathrm{s} =4$ in Vega) and $ K_\mathrm{s}
\le 21.1$ (AB). For EROs in our sample, we compute, using the
ACS F814W images, their concentration, asymmetry,
as well as their Gini coefficient and the second moment
of the brightest 20\% of their light.  Using those
morphology parameters and the $Spitzer$ IRAC $[3.6]-[8.0]$
color, the spectral energy distribution (SED) fitting method, we
classify EROs into two classes: old galaxies (OGs) and young,
dusty starburst galaxies (DGs). We found that the fraction of
OGs and DGs in our sample is similar, about 48 percentages of
EROs in our sample are OGs, and 52 percentages of them are DGs.
To reduce the redundancy of these three different classification
methods, we performed a principal component analysis 
on the measurements of EROs, and find that morphology parameters
and SEDs are efficient in segregating OGs and DGs. The $[3.6]-[8.0]$ color,
which depends on reddening, redshift, and photometric accuracy,
is difficult to separate EROs around the discriminating line
between starburst and elliptical.  We investigate the dependence
of the fraction of EROs on their observational properties, and
the results suggest that DGs become increasingly important
at fainter magnitudes, redder colors, and higher redshifts.
The clustering of the entire EROs, DGs, and OGs was estimated
by calculating their correlation function, and we
find that the clustering of EROs is much stronger than that of
full $K-$limited samples of galaxies; the clustering
amplitude of OGs is a factor of $\sim2$ larger than that of DGs.

\end{abstract}


\section{Introduction}

Understanding when and how the most massive galaxies in
the universe formed is one of the most outstanding problems
in cosmology and galaxy formation (Conselice et al. 2007).
At $z<1$, the most recent results seem now to agree in
indicating that the majority of massive galaxies were already
in place at z$\approx$ 0.7-0.8, with a number density consistent
with the one at $z=0$ (Yamada et al. 2005; Cimatti et al. 2006;
Bundy et al. 2007).  However, despite the remarkable success
in finding and studying massive galaxies over a wide range
of cosmic time, the global picture is far from being clear,
especially for massive galaxies at redshift $z\gtrsim1$
(Cimatti et al. 2008).

Deep imaging and spectroscopic surveys in the optical and 
near-infrared indicate that most of the massive galaxies at 
$z \gsim 1$ are extremely red objects (hereafter EROs, such as
Cimatti et al. 2004; Glazebrook et al. 2004; Conselice et al. 
2008).
Therefore, EROs are largely the test bed for galaxy models, 
and understanding their properties is an important test of the 
physics behind galaxy formation.  EROs, which are defined as 
objects having red optical-to-infrared colors, such as 
$I-K \ge 4$ or $ R-K \ge 5$ in the Vega-based magnitude 
system, were first discovered by deep near-infrared
surveys by Elston, Rieke, and Rieke (1988).  Since then many
authors have detected EROs in near-infrared surveys (McCarthy
et al. 1992; Thompson et al. 1999; Smith et al. 2002; 
Kong et al. 2006). Such very red colors can be produced 
by high-redshift ($z\gtrsim1$) old elliptical galaxies 
(hereafter OGs) with intrinsically red spectral energy 
distributions (SEDs), and can also be produced by dusty 
starburst galaxies whose UV luminosities are strongly 
absorbed by large amounts of internal dust (hereafter DGs; 
Hu \& Ridgway 1994; Graham \& Dey 1996; Smith et al. 2008).

EROs continue to attract considerable interest. On the one hand, 
the research in the literature suggests that they may well be 
the high-redshift counterparts and progenitors of local massive 
E and S0 galaxies. These two classes of EROs may represent 
different phases in the formation and evolution of high-redshift 
massive elliptical galaxies.  
The number densities of OGs provide the strongest constraints on 
models of galaxy evolution (Gonzalez-Perez et al. 2008).
On the other hand, DGs could be related to the ultraluminous
IR galaxies producing the bulk of the total energy in the
universe since the recombination era (Elbaz \& Cesarsky 2003).
The properties of EROs, specifically the fraction of OGs
and DGs, are therefore a crucial test of galaxy formation
theories, and will improve our ability to make complete
measurements of the star formation history of the universe
(Cimatti et al. 2004; Glazebrook et al. 2004; Renzini 2006).
Many groups are currently investigating the fractions of these
two ERO populations using a variety of observational approaches
(such as Moriondo et al. 2000; Stiavelli \& Treu 2001; 
Mannucci et al. 2002; Smail et al. 2002; Cimatti et al. 2003; 
Yan \& Thompson 2003; Giavalisco et al. 2004; McCarthy 2004; 
Moustakas et al. 2004; Sawicki et al. 2005; Doherty et al. 2005; 
Simpson et al. 2006; Conselice et al. 2008; Fang et al. 2009), 
but the fraction of OGs and DGs from different survey is different, 
even the same classified method was used. 
Therefore, their reliable classification and study, in particular 
the relative fraction of different ERO types, can provide crucial 
constraints on the evolution of massive, starburst, dusty, and/or 
ultraluminous infrared galaxies known to exist at higher redshift, 
e.g., BzK galaxies, distant red galaxies, and submillimeter and 
IR-luminous Lyman break galaxies.

Since EROs are rare and clustered, wide-field surveys are essential
for studies of the statistical properties of EROs (Daddi et al.
2000; Roche et al. 2002; Kong et al. 2006). Therefore, one
possibility of the difference among previous works is that the ERO
samples are small. The Cosmic Evolution Survey (COSMOS), the largest
contiguous survey ever with $Hubble$ $Space$ $Telescope$ ($HST$), provides 
for the first time a combined data set capable of simultaneously exploring 
large-scale structure and detailed galaxy properties (luminosity, size, color,
morphology, nuclear activity) out to a redshift approaching $z=1.5$.
The deep, panoramic multicolor data from Canada-France-Hawaii Telescope 
(CFHT)-$u^\star$, $i^\star$, Subaru Suprime-Cam $B_j$, $g^+$, $V_j$, $r^+$, $i^+$, 
and $z^+$, Cerro Tololo Inter-American Observatory (CTIO)/KPNO-Ks (hereafter $K$), 
$Spitzer$ Infrared Array Camera (IRAC; 3.6, 4.5, 5.8, 8.0 $\mu$m),
and the superb resolution $HST$ data enable us to classify the EROs
into OGs and DGs by different methods, such as the $[3.6]-[8.0]$
 infrared color (Wilson et al. 2007), the multi-wavelength SED 
fitting method, the morphological parameters of Gini coefficient ($G$), 
the second-order moment of the brightest 20\% of the galaxy's flux (\mt), 
concentration index ($C$), and rotational asymmetry ($A$). 
Considering the limited ERO sample in the previous papers, 
we will use those different methods to
classify $\sim$5300 EROs over the COSMOS filed in this paper. The
structure of this paper is as follows. We describe the COSMOS survey
in Section 2. The ERO sample is constructed in Section 3.
Classification of EROs, using the spectral type, infrared color,
morphological parameter, and the principal component analysis (PCA)
method, is performed in Section 4. In Section 5, we compare the
properties of OGs and DGs. A summary is given in Section 6.
Throughout this paper, we assume a standard cosmological model with 
$\Omega_M = 0.3$, $\Omega_\Lambda = 0.7$, and $h=H_{\rm0}$ 
(km s$^{-1}$ Mpc$^{-1}$)$/100=0.71$.

\section{The COSMOS Survey}

The COSMOS project is centered upon a complete survey in the
F814W band (central wavelength at 8332 \AA, with width 2510
\AA) using the Advanced Camera for Surveys (ACS) on board $HST$
of an area of 2 deg$^2$ (Scoville et al. 2007a). The 
COSMOS field is equatorial to ensure coverage by all astronomical 
facilities (centered at J2000.0 $\alpha$ = $10^{\rm h} ~ 
00^{\rm m} ~ 28^{\rm s}.6$, $\delta$ = $+02^\circ ~ 12' ~ 21''.0$), 
which was chosen to be devoid of bright X-ray, UV, and radio sources. 
An overview of the COSMOS project is given in Scoville et al. (2007b).

With an ACS field of view of 202 arcsec on a side, this required
a mosaic of 575 tiles, corresponding to one orbit each, split
into four exposures of 507 s dithered in a four-point line
pattern. The final images have absolute astrometric accuracy
of better than 0.1 arcsec. A version with 0.05 arcsec
pixels was used to measure the morphology of galaxies in the
structure. More details and a full description of the ACS data
processing are provided in Koekemoer et al. (2007).

Ground-based follow-up observations have been performed
using the CFHT Megacam ($u^\star$ and $i^\star$), Subaru
Suprime-Cam ($B_j$, $g^+$, $V_j$, $r^+$, $i^+$, and $z^+$),
Kitt Peak Flamingos ($K$ band), and CTIO (also $K$ band) telescopes,
providing deep coverage, with typical limiting magnitudes
of 27 (AB, $3\sigma$), of the field from the $u^\star$ to
$z^+$ bands, as well as shallower imaging in the $K = 21.6$
(AB). Details of the ground-based observations and data
reduction are presented in Capak et al. (2007a) and Taniguchi
et al. (2007). A multi-wavelength photometric catalog (Capak et
al. 2007a) was generated using SExtractor (Bertin \& Arnouts
1996), with the $i^+$ band as the selection wavelength.

In additional, a deep infrared imaging survey for the COSMOS
field (S-COSMOS) was carried out with the $Spitzer$ $Space$
$Telescope$ (Werner et al. 2004). 166 hours of observations with
the $Spitzer$ IRAC camera (Fazio et al. 2004) have been dedicated
to cover the entire 2 deg$^2$ COSMOS field. The field has
been observed simultaneously in the four IRAC channels$-$3.6,
4.5, 5.8, and 8.0 $\mu$m. 54 hours of observations with the
Multiband Imaging Photometer for Spitzer (MIPS) camera 
(Rieke et al. 2004) have been dedicated to cover
the entire 2 deg$^2$ COSMOS field with a shallow survey
(16 hr) at 24, 70, 160 $\mu$m, and a small "test" area
(0.16 deg$^2$) with very deep observations (38 hr)
at 24, 70, 160 $\mu$m in the Cycle 2. More details and a
full description of the $Spitzer$ data processing are provided
in Sanders et al. (2007).

\section{EROs Selection}

Compared to optical, the near-infrared selection (in particular
in the $K$ band) offers several advantages, including the
relative insensitivity of the $k$-corrections to the galaxy type
even at high redshift, the less severe dust extinction effects,
the weaker dependence on the instantaneous star formation
activity, and a tighter correlation with the stellar mass of
the galaxies. Therefore, the studies of faint galaxy samples
selected in the near-infrared have long been recognized as ideal
tools to study the process of mass assembly at high redshift
(Broadhurst et al. 1992; Kauffmann \& Charlot 1998; 
Cimatti et al. 2002; Bundy et al. 2006). Therefore,
we selected objects in the COSMOS survey using the K-band
limiting magnitude.

The limiting magnitude (in AB) of the $K-$band image was
defined as the brightness corresponding to 5$\sigma$ on a
3$''$ diameter aperture for an isolated point source by Capak
et al. (2007a), the value is $K = 21.6$ (AB). The edges of the
$K-$band image have low signal-to-noise value, therefore, the
area, as discussed in this paper, was reduced from the COSMOS
2 deg$^2$ to 1.82 deg$^2$. 3234,836 objects were
detected and included in the Capak et al.'s {\bf ALL} catalog
(v20060103), and 2797,708 of them are in the 1.82 deg$^2$
area (Capak et al. 2007a). We selected objects to $\Kv<19.2$
($K$-band total magnitude, $\sim 21.1$ in AB), over a total
sky area of 1.82 deg$^2$. The total magnitudes were
defined as the brightest between the Kron automatic aperture
magnitudes and the corrected aperture magnitude. The aperture
corrections were estimated from the difference between the
Kron automatic aperture magnitudes and the 3$\arcsec$ aperture
magnitudes. Simulations of point sources show that in all the
area the completeness is well above 90\% at this K-band level.

\begin{figure}
\centering
\includegraphics[angle=-90,width=\columnwidth]{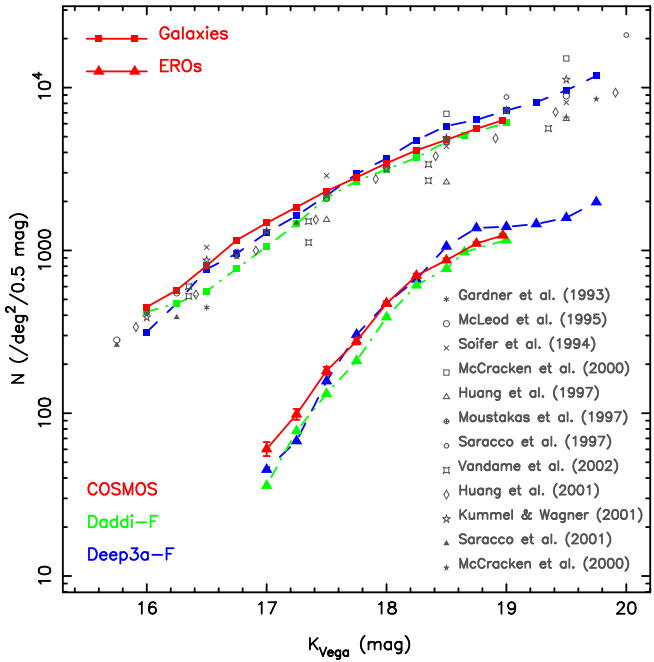}
\caption{
$K$-band differential number counts for field galaxies and
EROs in COSMOS, compared with a compilation of results taken
from various sources. The filled squares correspond to field
galaxies, and triangles correspond to EROs. Red, green, and
blue color correspond to the counts of galaxies and EROs in
the COSMOS (this paper), Daddi-F and Deep3a-F (Kong et al.
2006), respectively. The error bars of EROs indicate the 
Poissonian uncertainties.
}
\label{fig:numc}
\end{figure}

Therefore, we selected objects to $\Kv<19.2$, over a total sky
area of 1.82 deg$^2$, and 34,391 objects are included
in our final catalog. Stellar objects are isolated with
the color criterion $(z-K)_{\rm AB}<0.3(B-z)_{\rm AB}-0.5$
(Daddi et al. 2004), same as in Kong et al. (2006). 626 objects
were classified as stars, and the number of the final galaxy
sample is 33,765. Figure~\ref{fig:numc} shows a comparison of
K-band number counts in the COSMOS survey with a compilation
of literature counts. The red-, green-, and blue-filled
squares correspond to the counts of field galaxies in the
COSMOS (this paper), Daddi-F, and Deep3a-F (Kong et al. 2006),
respectively. As shown in the figure, our counts are in good
agreement with those of previous surveys.

\begin{figure*}
\centering
\includegraphics[angle=-90,width=\textwidth]{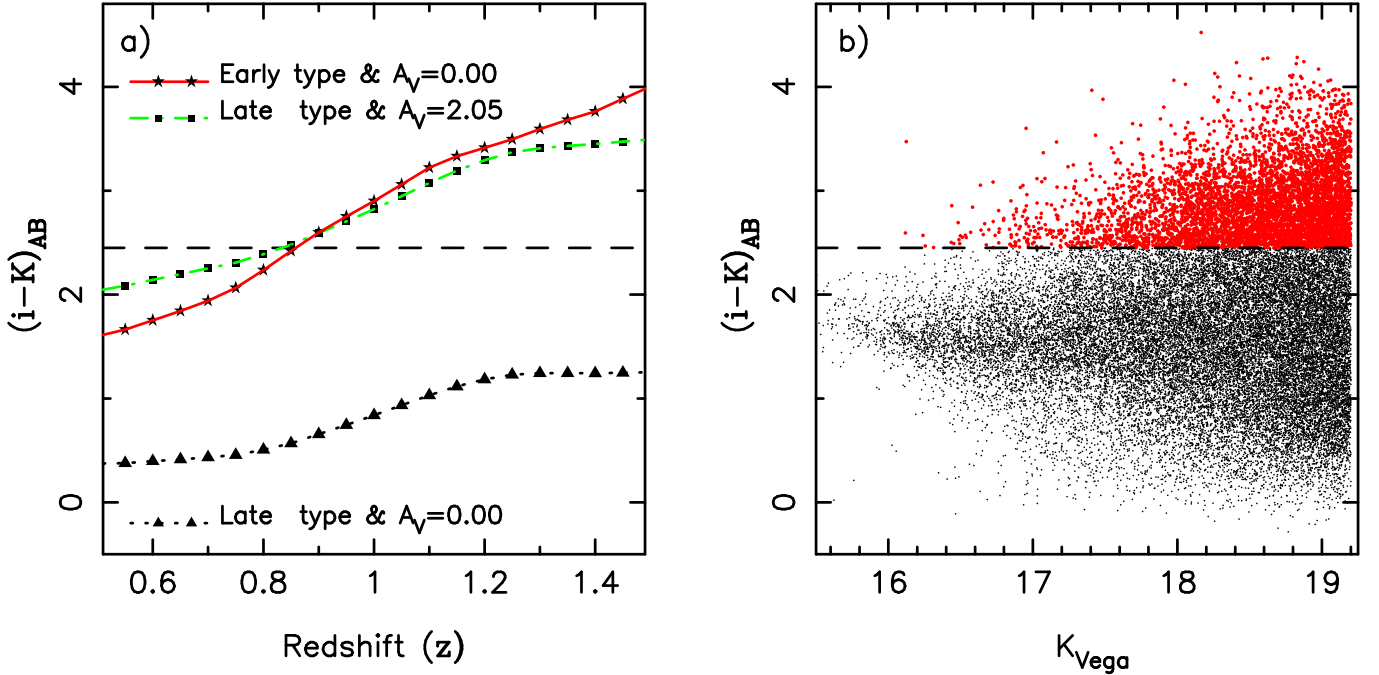}
\caption{
(a) Model $i-K$ color as a function of redshift. Representative 
model galaxy SEDs are taken from Kodama \& Arimoto's (1997; KA97)
 population synthesis models: a late-type galaxy with 
$A_{\rm V}=0.0$ (triangles), 
an early-type galaxy with $A_{\rm V}=0.0$ (stars), 
and a late-type galaxy with $A_{\rm V}=2.05$ (squares). 
(b) $i-K$ plotted against $K$ for all galaxies in the $K$-limited
sample. Red circles denote EROs.
The horizontal dashed line corresponds to the color threshold 
($i-K = 2.45$ in AB) for selecting EROs in this paper.}
\label{fig:ik-k}
\end{figure*}

Figure ~\ref{fig:ik-k}(a) shows $i-K$ ($i$ is the Subaru 
Suprime-Cam $i^+$ filter) model color of several 
representative galaxies against redshift.
Model SEDs are adopted from the Kodama \& Arimoto's (1997; KA97)
 population synthesis library. 
Ordinary late-type galaxies never reach red color, while both 
early-type galaxies and reddened late-type galaxies cross the 
$i-K=2.45$ line when seen beyond $z \sim 0.8$.
Therefore, in this paper we define EROs as objects whose $i-K$ 
 color is equal to or redder than 2.45 in AB, which is similar 
to $I-K=4.0$ in Vega.  
Figure~\ref{fig:ik-k}(b) plots $i-K$ as a function of the $\Kv$ 
magnitude for all galaxies in our sample. The horizontal line 
denotes our boundary for the ERO selection. 
There are 5264 objects which satisfy the ERO threshold, 
$i-K \ge 2.45$, down to $\Kv = 19.2$ mag. We refer to 
these objects as the ERO sample.

The $K$-band differential number
counts of the EROs in the COSMOS field (red triangles) are shown
in Figure.~\ref{fig:numc},
together with those in Deep3a-F (blue triangles) and Daddi-F
(green triangles; Kong et al. 2006). From this figure, we found
that the counts of COSMOS match very well with those of Deep3a-F
and Daddi-F. It can be explained easily, since the areas of 
those three fields are large (from $320$ arcmin$^2$ to 1.82 deg$^2$),
 the field-to-field variation in the number counts is very small.
Another feature seen in previous surveys, which we also see, 
is a turnover in the slope of the counts at $\Kv=18.5$, being
steeper at bright magnitudes and flattening out toward faint
magnitudes. 
As we saw in Figure~\ref{fig:ik-k}(b), and will see in 
Figure~\ref{fig:photoz}(b), $i-K$ color cut can effectively remove 
the $z < 0.8$ foreground from galaxy surveys. The steep slope of 
the counts at relatively bright magnitudes reflects the 
exponential cutoff of the galaxy luminosity function. Unlike the 
full $K$-selected counts, the red color-selected objects span a 
fairly narrow redshift range, and thus the shape of the counts 
closely reflects the shape of the luminosity function, i.e., most
of the EROs are luminous galaxies.

Then we match our ERO sample with the S-COSMOS data (Sanders
et al.  2007). The catalog of S-COSMOS IRAC GO2 Delivery
(v27May2007) includes photometry in the four IRAC channels for
all those sources that have a measured flux in IRAC Channel
1 (3.6 $\mu$m above 1 $\mu$Jy), 345,512 objects were included. 
We match
the 5264 EROs in our sample with the S-COSMOS IRAC catalog,
using a 2$''.5$ radius around the Subaru $i^+$-band position,
and 5255 EROs have IRAC counterparts, nine EROs have not IRAC
counterparts. We also match our ERO sample with the S-COSMOS
MIPS-Ge GO2 Delivery (v27May2007), and found 325 of them have
counterparts in the MIPS 24 $\mu$m catalog.

\section{Classification of EROs}

Since EROs can be divided into two broad classes, passively
evolving OGs and DGs, the populations should be differentiated 
before we attempt to analyze their properties.  In this section, 
we shall classify our EROs into two classes, using three different 
methods: SED fitting, infrared color, and morphology. In the 
last subsection, we also perform the PCA method to our EROs sample, 
and discuss the differences.

\subsection{Classification based on SED Fitting}\label{sec:sed}

An SED fitting technique based on an updated version of the
$Hyperz$ code (Bolzonella et al. 2000) is used to classify
our ERO sample into different type, dusty and evolved using
their multi-waveband photometric properties. This technique has
been used in many previous works, such as Smail et al. (2002),
Miyazaki et al. (2003), Georgakakis et al. (2006) 
and Fang et al. (2009).

We use a stellar population synthesis model by KA97 to make
template SEDs. KA97 include the chemical evolution of gas and
stellar populations, and have been successfully used to obtain
photometric redshifts of high- and low-redshift galaxies (Kodama
et al. 1999, Furusawa et al. 2000). The template SEDs consist
of the spectra of pure disks, pure bulges, and composites made
by interpolating the two as shown in Table~\ref{tab:sed}. Pure
disk SEDs correspond to young or active star-forming galaxies,
and pure bulge SEDs correspond to elliptical galaxies. 
Model parameters for the bulge galaxies are calibrated to 
reproduce the average color of elliptical galaxies in clusters 
of galaxies, and for the disk galaxies, which are close to the 
values estimated for the disk of our Galaxy.  

The intermediate SED types were made by combining a disk 
component and a bulge component with the same age, but with 
different star formation histories. The ratio of the
bulge luminosity to the total luminosity in the B band, which
we define as $f_{old}$, is changed from 0.1 to 0.99, as shown
in the second line of Table~\ref{tab:sed}.  $N_{age}$ is the
number of the template SED age, 33 different ages from 0.01
Gyr to 15 Gyr for pure disk templates, and 15 different ages
from 1 Gyr to 15 Gyr for others. In total, our basic template
set consists of $15\times12+33=213$ SEDs. More details about
the templates can be found in Furusawa et al. (2000).

\begin{table*}
\centering
\scriptsize
\caption{Template SEDs from 
the stellar population synthesis model by Kodama \& Arimoto 
(1997) adopted in this work.}
\label{tab:sed}
\vspace{0.2cm}
\begin{tabular*}{1.0\textwidth}{@{\extracolsep{\fill}}l|crrrrrrrrrrrc}
\tableline
\tableline
Type& \multicolumn{1}{c}{Pure Disk} &
\multicolumn{11}{c}{Bulge+Disk\tablenotemark{a}}
&\multicolumn{1}{c}{Pure Bulge}\\
\cline{1-1} \cline{2-2} \cline{3-13} \cline{14-14}
$f_{old}$\tablenotemark{b} &
0.0&0.1&0.2&0.3&0.4&0.5&0.6&0.7&0.8&0.9&0.95&0.99&1.0 \\
$N_{age}$\tablenotemark{c} &  33& 15& 15& 15& 15& 15& 15& 15&
15& 15&  15&  15&15  \\
$\SpT$\tablenotemark{d}       &   1&  2&  3&  4&  5&  6&  7&
8&  9& 10&  11&  12&13  \\
\tableline
\end{tabular*}
\tablenotetext{a}{Bulge+disk means a combination of a passive
(pure bulge-like) spectrum and a star-forming (pure disk-like) 
spectrum.}
\tablenotetext{b}{$f_{old}$ is the fraction of the bulge 
(passive) component luminosity in the total luminosity at the $B$ band 
of a template.}
\tablenotetext{c}{$N_{age}$ is the number of the template age,
33 different ages
for pure disk templates, and 15 different ages for others.}
\tablenotetext{d}{$\SpT$ is the number of the spectral 
type of templates, 1 = pure disk (young), 13 = pure bulge 
(old).}
\end{table*}

The SED derived from the observed
magnitudes of each object is compared to each template spectrum
(redshift from 0.0 to 6.5 with step 0.05; $A_{\rm V}$ from 0.0
to 6.0 with step 0.05; Calzetti et al.'s internal reddening
law, Calzetti et al. 2000) in turn. The weighted mean redshift,
computed in the confidence intervals at 99\% around the main
solution, from $Hyperz$, was calculated for each object. For
all 5264 EROs in our sample, 5255 of them have four IRAC channel
magnitudes, the 13 band magnitudes, including CFHT-$u^\star$ and
$i^\star$, Subaru-$B_j$, $g^+$, $V_j$, $r^+$, $i^+$, and $z^+$,
near-infrared $K$ band, and IRAC 3.6, 4.5, 5.8 and 8.0 $\mu$m
bands, were used to calculate photometric redshift and classify
EROs. Nine of them have not IRAC counterparts, only optical-
and near-infrared $K-$band data were used for calculations.

\begin{figure*}
\centering
\includegraphics[angle=-90, width=\textwidth]{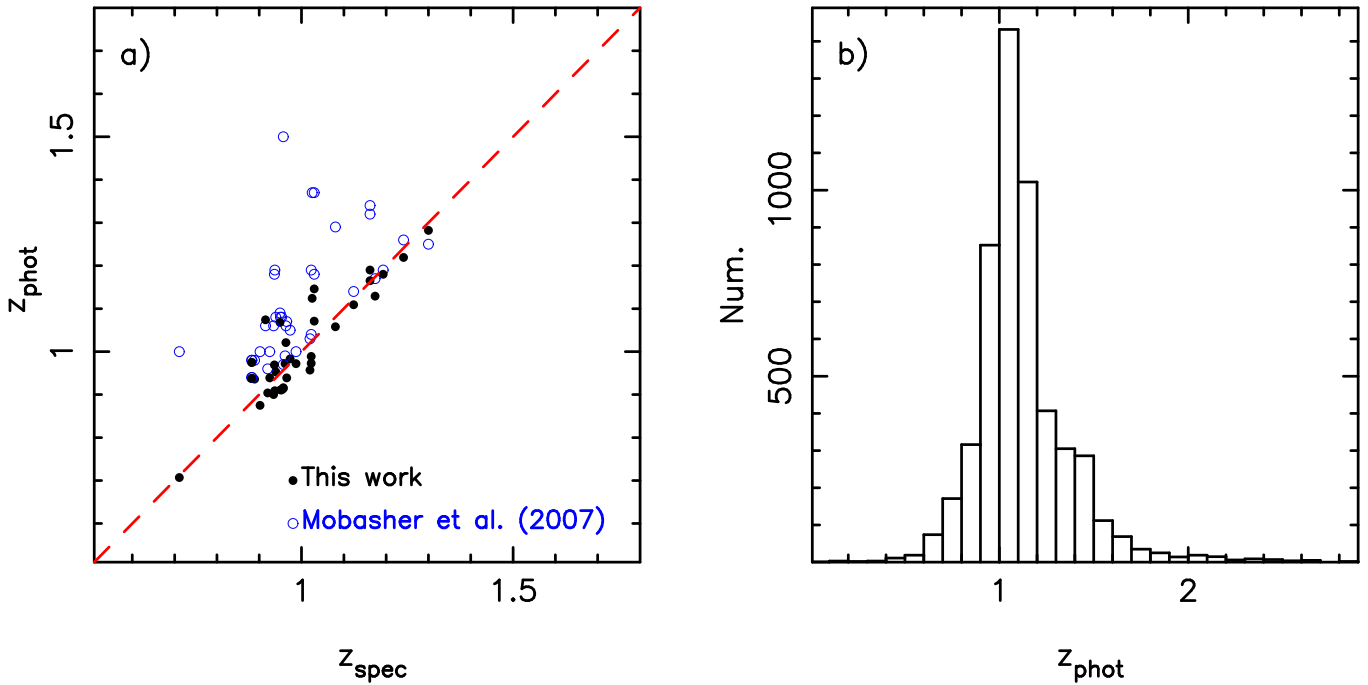}
\caption{
(a) Comparison between photometric and spectroscopic redshifts
for a sample of 38 EROs in COSMOS with available spectroscopic
redshifts.  The filled circles correspond to photometric redshifts
from this work, and open circles correspond to photometric
redshifts from Mobasher et al. (2007). (b) Photometric redshift
distribution of EROs in the COSMOS field.
}
\label{fig:photoz}
\end{figure*}

Thirty-eight EROs in our sample have spectroscopic redshifts 
($0.71<z<1.30$) from 
zCOSMOS (Lilly et al. 2007). In Figure~\ref{fig:photoz}(a), we 
show a comparison of the photometric redshift from our SED 
fitting method with the spectroscopic redshift for those EROs;
the range of spectrometric redshifts does not extend to lower
or higher redshift. 
In additional, we plot the photometric
redshift from Mobasher et al. (2007) in this figure as well,
which were calculated with optical and near-infrared bands,
and E/S0, Sab, Sc, Im, and two starburst templates 
(Ilbert et al. (2009) gave the photometric redshift of 
$i^+ < 26$ galaxies 
in the COSMOS field with 30 band images, but we do not get 
the data yet). 
The filled and open circles show objects where the photometric
redshift are taken from our method and from Mobasher et al. (2007),
respectively. From this figure, we found that our photometric
redshifts fit the spectroscopic redshifts well, with an average
$\delta z/(1+z_{spec})=0.02$. Photometric redshifts of EROs 
from Mobasher et al. (2007) were systemic overestimated, 
with an average $\delta z/(1+z_{spec})= 0.06$. 
The main reason for this difference is that the IRAC four band 
data were not used in Mobasher et al. (2007). In addition, 
the templates used in this work are different from 
those of Mobasher et al. (2007), where four normal galaxy 
templates (E, Sbc, Scd, and Im) from Coleman et al. (1980) 
and two starburst templates (SB2 and SB3) from Kinney et al. 
(1996) were used.

We plot the photometric redshift histogram of EROs in our
sample in Figure~\ref{fig:photoz}(b). The
average redshift and the peak of the redshift distribution
of EROs in our sample are at $z_{phot} \sim 1.1$, only less
than 5\% of them have $z_{phot}<0.8$. The photometric redshift
distribution supports that the color criteria $i-K \geq 2.45$
is valid for culling galaxies at $z\gsim 0.8$, and that
the redshift estimates by the spectrum fitting method are
reasonable. We have checked the effect from templates, with a
set of synthetic templates obtained using the GALAXEV03 code
(Bruzual \& Charlot 2003), and found similar results from
those different templates.

\begin{figure*}
\centering
\includegraphics[angle=-90,width=\textwidth]{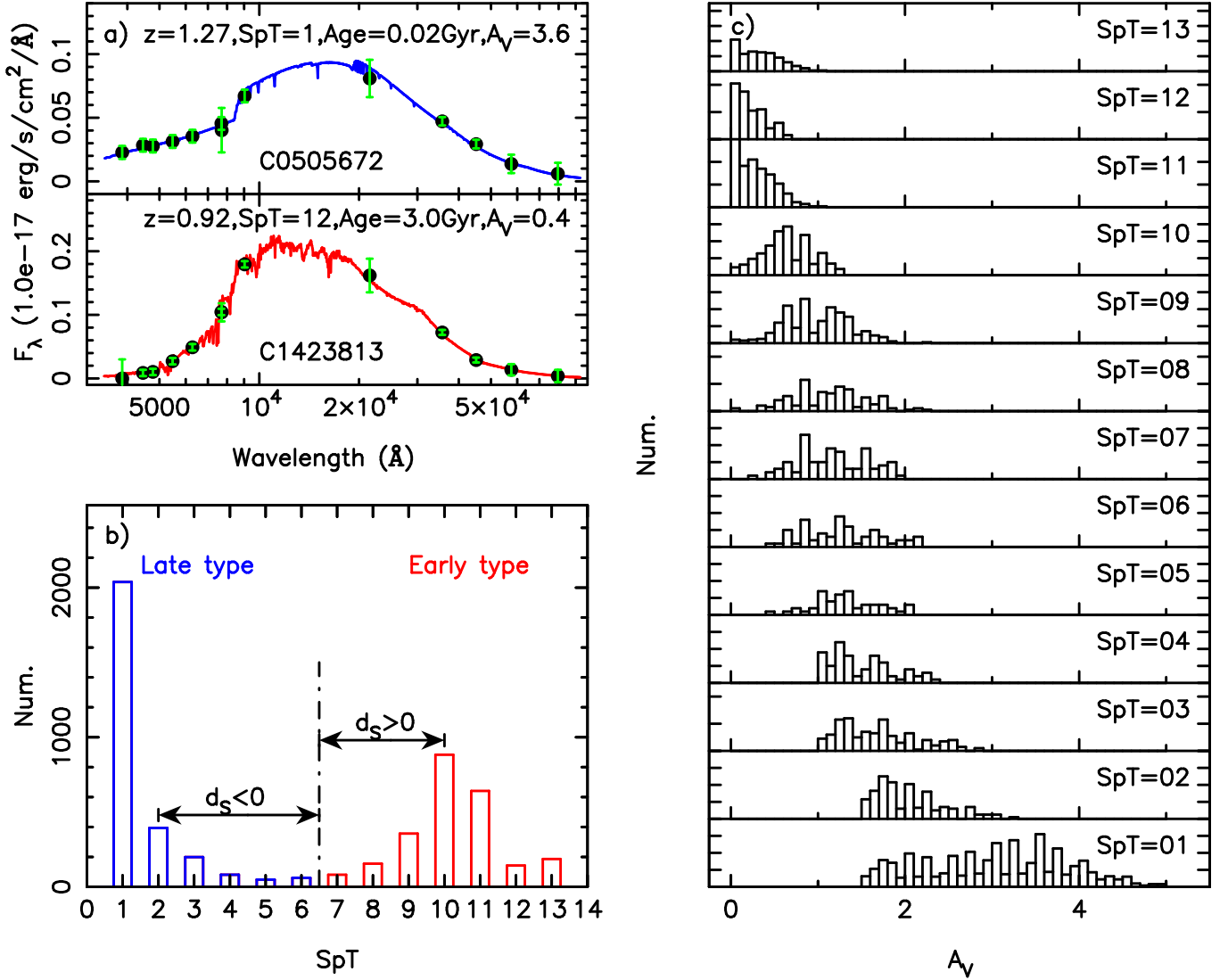}
\caption{
(a) SED of two EROs in the COSMOS 
field. The filled circles show the observed photometric data,
and the solid curves show the best-fitting templates. Photometric
redshift, number of the spectral type, age, and dust extinction
in the $V$ band for each EROs are also shown. 
(b) Distribution of spectral type (\SpT) for EROs 
in the COSMOS field. $\SpT=1$: pure disk (late type); $\SpT=13$: 
pure bulge (early type). $\dS=\SpT-6.5$.
(c) Distribution of dust extinction in the $V$ band ($A_{\rm V}$)
for EROs in our sample.
}
\label{fig:spt}
\end{figure*}

Given as an example, we show SED of
two EROs in our sample, with the best-fitting template from
KA97 as shown in Figure~\ref{fig:spt}(a). Some outputs from $Hyperz$, 
such as photometric redshift, the number of the spectral type 
(see the last line of Table~\ref{tab:sed}), age, and dust 
extinction in the $V$ band, for each ERO are also shown. 
The top panel of Figure~\ref{fig:spt}(a) shows an ERO, which 
SED can be fitted with a late-type template, the age of 
stellar population is young (0.02 Gyr), and the dust 
extinction is high ($A_V=3.6$), so it is a DG with redshift $z=1.27$. 
The bottom panel of Figure~\ref{fig:spt}(a) shows another
EROs, which SED can be fitted with an early-type template,
the age of stellar population is old (3 Gyr), and the dust 
extinction is low ($A_V=0.4$), it is an OG with redshift 
$z=0.92$.

In Figure~\ref{fig:spt}(b), we show the number of the 
spectral type ($\SpT$) distribution for our ERO 
sample, we found that most of the EROs can be fitted well by the 
templates with $\SpT\leq3$ (disk-dominant galaxies, DGs) or 
with $\SpT\geq8$ (bulge-dominant galaxies, OGs). Only 5\% of 
EROs in our sample are well fitted with intermediate templates 
($\SpT$ from 4 to 7), because the number is small, they have 
very limited effect to ERO's classification. 
Based on these analysis, we classify EROs as DGs, if the $\SpT$ 
of the best-fitting template less than 6.5 ($\SpT < 6.5$, the 
ratio of the bulge luminosity to the total luminosity in the B 
band less than 50\%), the others were classified as OGs. 
Out of the 5264 EROs in our sample, 2505 ($\sim 48\%$) are 
classified as OGs, while 2759 ($\sim 52\%$) are classified as 
DGs. Therefore, based on the SED fitting classification method, 
we concluded that the fraction of OGs and DGs in the COSMOS 
field is similar.

Figure~\ref{fig:spt}(c) shows the distribution of dust extinction 
in the $V$ band ($A_{\rm V}$) for EROs in our sample. 
EROs with a large $\SpT$ (bulge-dominant galaxies) are found to 
have small $A_{\rm V}$ values, which is fully consistent with the
properties of passively evolving ellipticals with small amounts 
of dust. 
DGs with $\SpT \leq 2$ are found to have 
$ A_{\rm V} \sim 1.5 - 4.5$, with a median value of $ \sim 3.1$. 
This corresponds to  $E(B-V)\sim 0.77$ (Calzetti's extinction law
was adopted). 
The $A_{\rm V}$ distribution of our DGs is in good agreement with 
that of Cimatti et al.'s (2002). 
They have found that the average spectrum of DGs for which they 
made spectroscopy is fitted by a local starburst galaxy with 
$ E(B-V) \sim 0.8$. They have also found that the global shape 
of the continuum and the average $R-K$ color of their DGs can 
be reproduced by synthetic spectra of star-forming galaxies 
with $ 0.6 < E(B-V) < 1.1$.

\subsection{Classification based on $[3.6]-[8.0]$ Color}

Since the COSMOS survey does not release the $J$-band data
yet, so we cannot classify EROs into OGs and DGs based on
their locations in the $J-K$ versus $R-K$ plane (Pozzetti \&
Mannucci 2000). However, in the rest-frame near-infrared,
old stellar populations show a turndown at wavelengths longer
than the rest-frame 1.6 $\mu$m "bump", while dusty starburst
populations show emission from small hot dust grains, therefore,
the near- and mid-infrared data from $Spitzer$ can be used to
help us distinguish among different ERO populations. Wilson et
al. (2007) introduced a new method, based on IRAC $[3.6]-[8.0]$
color, and classified EROs as OGs, DGs, or active galactic
nuclei (AGNs).  They found that OGs are expected to have the
bluest colors ($[3.6] - [8.0] \sim -1.0$), DGs are expected
to have intermediate colors ($[3.6] - [8.0] \sim -0.5$),
and AGNs are expected to have the reddest colors ($[3.6] -
[8.0] \sim1.0$). $[3.6] - [8.0] < -0.75$ was used to classify
EROs as OGs, $-0.75 < [3.6] - [8.0] < 0.0$ was used to classify
EROs as DGs, and galaxies with $[3.6] - [8.0] > 0.0$ were used
to classify EROs as AGNs.

\begin{figure*}
\centering
\includegraphics[angle=-90,width=\textwidth]{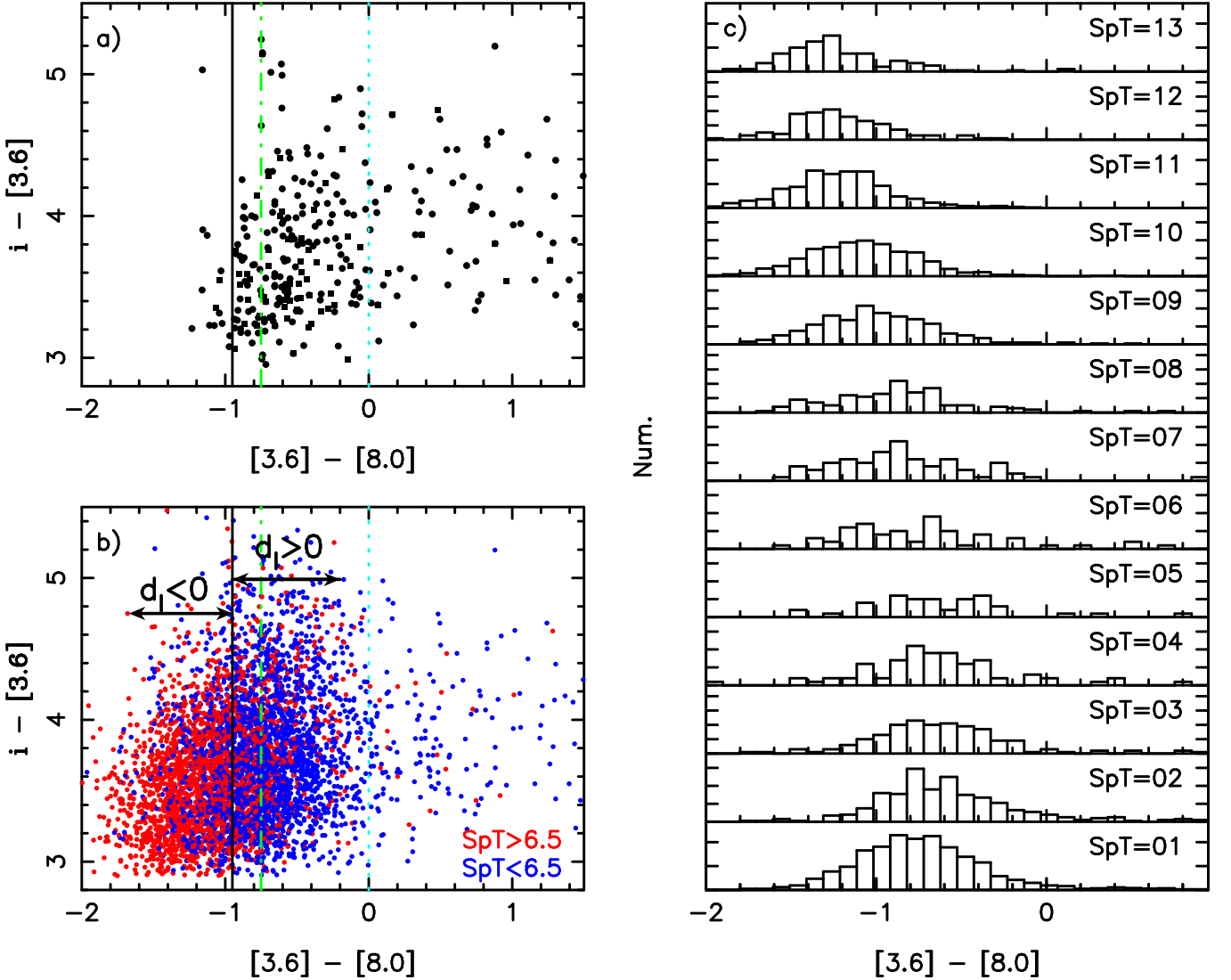}
\caption{
$i-[3.6]$ vs. $[3.6]-[8.0]$ color--color diagram. 
(a) 325 EROs with MIPS 24 $\mu$m detection (DGs);
(b) 4761 EROs in our sample with the IRAC 3.6 and 8.0 $\mu$m 
detection. The dotted line and dot-dashed line show the color 
criteria were used in Wilson et al. (2007), to classify EROs 
into OGs, DGs, and AGNs. 
The solid line, $[3.6]-[8.0]=-0.95$, was used to classify 
EROs into OGs and DGs in our works. Red symbols indicate 
EROs with $\SpT > 6.5$, and were classified as OGs with the SED 
fitting method; and blue symbols are EROs with $\SpT < 6.5$.  
$\dI=[3.6]-[8.0]-(-0.95)$ is the distance from an object at 
$i - [3.6]$ vs. $[3.6]-[8.0]$ diagram to the 
$[3.6]-[8.0]=-0.95$ line. 
(c) $[3.6]-[8.0]$ distribution for EROs with different $\SpT$.
}
\label{fig:irac}
\end{figure*}

Strong, rest-frame $6 \sim 12 \mu$m polycyclic aromatic
hydrocarbon (PAH) and very small dust grains features redshift
into the 24 $\mu$m band at $z\sim 1 -2$, DGs are clearly
distinguished from OGs at mid-infrared (Yan et al. 2004;
Stern et al. 2006).  
Because the very deep MIPS observation for the whole COSMOS 
field is not yet finished, Figure~\ref{fig:irac}(a) shows the 325 
EROs in our sample, which were detected on the shallow MIPS 
24 $\mu$m images (Sanders et al. 2007). They are bright DGs, 
so can be detected on the shallow MIPS 24 $\mu$m image. 
From this figure, we found that most of them have 
$[3.6]-[8.0]>-0.95$, the color $[3.6]-[8.0]=-0.75$ from Wilson
 et al. (2007) is too red to be used to classify OGs and DGs. 
This can be confirmed with Figure 1 (the evolution track 
of early-type template) and Figure 2 (EROs with MIPS 24 $\mu$m
detection) of Wilson et al. (2007) as well. Therefore, we will
use $[3.6]-[8.0]=-0.95$ as the color criterion to classify
EROs as OGs and DGs in this paper.

We have a total of 5264 EROs in our sample, 5255 of them
are detected at 3.6 $\mu$m, but only 4761 of them with
fluxes brighter than the $3\sigma$ flux limit of 1 $\mu$Jy
at both 3.6 and 8.0 $\mu$m band images. We will apply the
$[3.6]-[8.0]=-0.95$ color criterion to classify these 4761
EROs in our sample. 
Figure~\ref{fig:irac}(b)
shows the distribution of EROs in the $[3.6]-[8.0]$ versus the 
$i-[3.6]$ plane. Two thousand one hundred and seventy-six of 
them (45\%) have $[3.6]-[8.0]<-0.95$,
which can be classified as OGs; 2494 of them (51\%) have
$-0.95<[3.6]-[8.0]<0.0$, which can be classified as DGs, and $\sim$4\%
of them with $[3.6]-[8.0]>0.0$ are AGNs. The fraction of AGN
among the ERO population here is similar to that in Brusa
et al. (2005). Using an 80 ks $XMM-Newton$ observation of a
sample of near-infrared selected EROs ($\Kv\le19.2$; Daddi
et al. 2000), the authors found that the fraction of AGN EROs
within near-infrared-selected ERO samples is $\sim$3.5\%.

In Section ~\ref{sec:sed}, we have classified EROs as OGs and DGs,
based on the SED fitting method. EROs with $\SpT<6.5$ are DGs, and
with $\SpT>6.5$ are OGs. We colored the symbols with the $\SpT$ 
of EROs in Figure~\ref{fig:irac}(b), red indicates OGs with 
$\SpT>6.5$, and blue indicates DGs with $\SpT<6.5$.  
We found that most of the EROs with $\SpT>6.5$ sit on the area with 
$[3.6]-[8.0]<-0.95$; however, EROs with $\SpT<6.5$ sit on the 
area with $[3.6]-[8.0]>-0.95$. 

To show this point more clearly, we plot in 
Figure~\ref{fig:irac}(c) the $[3.6]-[8.0]$ distribution
of EROs in our sample, from $\SpT=1$ (pure disk) to $\SpT=13$ 
(pure bulge), which were classified by the SED fitting method. 
We found that, in general, EROs with small $\SpT$ number (DGs) 
have red $[3.6]-[8.0]$ color (most of them $>-0.95$), and with 
large $\SpT$ number (OGs) have blue $[3.6]-[8.0]$ color.
However, the $[3.6]-[8.0]$ distribution shows that there is a 
significant overlap among EROs with different $\SpT$.
As we shall discuss in Section 4.4, the $[3.6]-[8.0]$ color cut
cannot separate EROs very well, which depends on reddening, 
redshift, and photometric accuracy.

\subsection{Classification based on Morphology}

As different morphologies are expected in the case of OGs (early
type) or DGs (late type), $HST$ imaging can be used to classify
EROs. Visual morphological classification on nearby galaxies
has had a significant impact on our understanding of galaxy
formation, environment, and evolution. However, it is widely
acknowledged that visual classification is an inherently
uncertain and subjective process. In the high-redshift regime,
the visual classification of galaxy morphology is further
complicated by limited resolution (even with $HST$). Another
method is to describe a galaxy parametrically, by modeling the
distribution of light as projected into the plane of the sky
with a prescribed analytic function, but they assume that the
galaxy is well described by a smooth, symmetric profile$-$an
assumption that breaks down for irregular, tidally disturbed,
and merging galaxies (Peng et al. 2002; Simard et al. 2002;
Balcells et al. 2003).

Nonparametric measures of galaxy morphology, such as Gini
coefficient (the relative distribution of the galaxy pixel flux
values, or $G$), \mt\, (the second-order moment of the brightest
20\% of the galaxy's flux), $C$ (concentration index), and $A$
(asymmetry), do not assume a particular analytic function for
the galaxy's light distribution and therefore can be applied to
irregulars, as well as standard Hubble-type galaxies (Gini 1912;
Abraham et al. 1996, 2003; Bershady et al. 2000; Conselice 2003;
Lotz et al. 2004). To classify EROs as OGs or DGs, we have
measured these nonparametric quantification ($G$, \mt, $C$,
and $A$) of galaxy structure for 4455 EROs in our sample with
the $HST$ ACS F814W images. Specific details of these measurements
can be found in Abraham et al. (2007). For the other 809 EROs,
we cannot measure these nonparametric quantifications for them,
since they are located at the edge of our study area, and have no
$HST$/ACS observation, or the signal-to-noise of $HST$/ACS images
is low.

\begin{figure}
\centering
\includegraphics[angle=-90,width=\columnwidth]{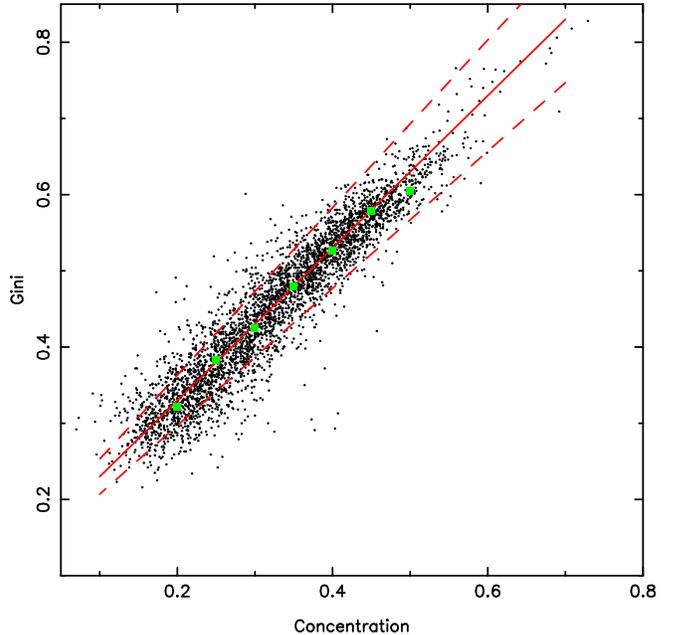}
\caption{
Gini coefficient ($G$) vs. central concentration ($C$) for EROs.
The solid line corresponds to unity slope, the big squares 
show the median value of $G$ at different $C$ bins, from 0.2 to
0.55. The dashed lines are $\pm10$\% offset relative to the 
solid line.
}
\label{fig:gc}
\end{figure}

Figure~\ref{fig:gc} illustrates the relationship between
the Gini coefficient and the central concentration for EROs
in our sample. The big squares represent the median values 
of $G$ at different $C$ bins. The solid line in 
Figure~\ref{fig:gc} corresponds to a slope of unity and a 
$y-$axis intercept of 0.13 (i.e., a given galaxy has a 
slightly greater value of $G$ than $C$). 
The dashed lines show a $\pm10$\% offset relative to 
the solid line.
EROs in our sample span a broad range of morphologies, from 
pure disk systems at low $C$ to highly centrally concentrated 
elliptical galaxies at high $C$.
Although there is clearly a very strong correlation between 
$G$ and $C$, but the $C$-$G$ relation of EROs exhibits
large scatter, objects with low central concentration (late-type 
galaxies) exhibit greater scatter. The reason for the
scatter is that measurements of central concentration have
been based on simple aperture photometry, therefore rely on
two key assumptions. First, the measurements depend on an
assumed symmetry in the galaxies. The second assumption is
that galaxy images have a well-defined center. Inspection
of the images of high-redshift galaxies shows that neither
of these assumptions is likely to be fulfilled when studying
galaxies in the distant universe (Abraham et al. 2003; Lotz
et al. 2004). Therefore, we will not use concentration index
as a classification parameter for EROs in this paper.

\begin{figure*}
\centering
\includegraphics[angle=-90,width=\textwidth]{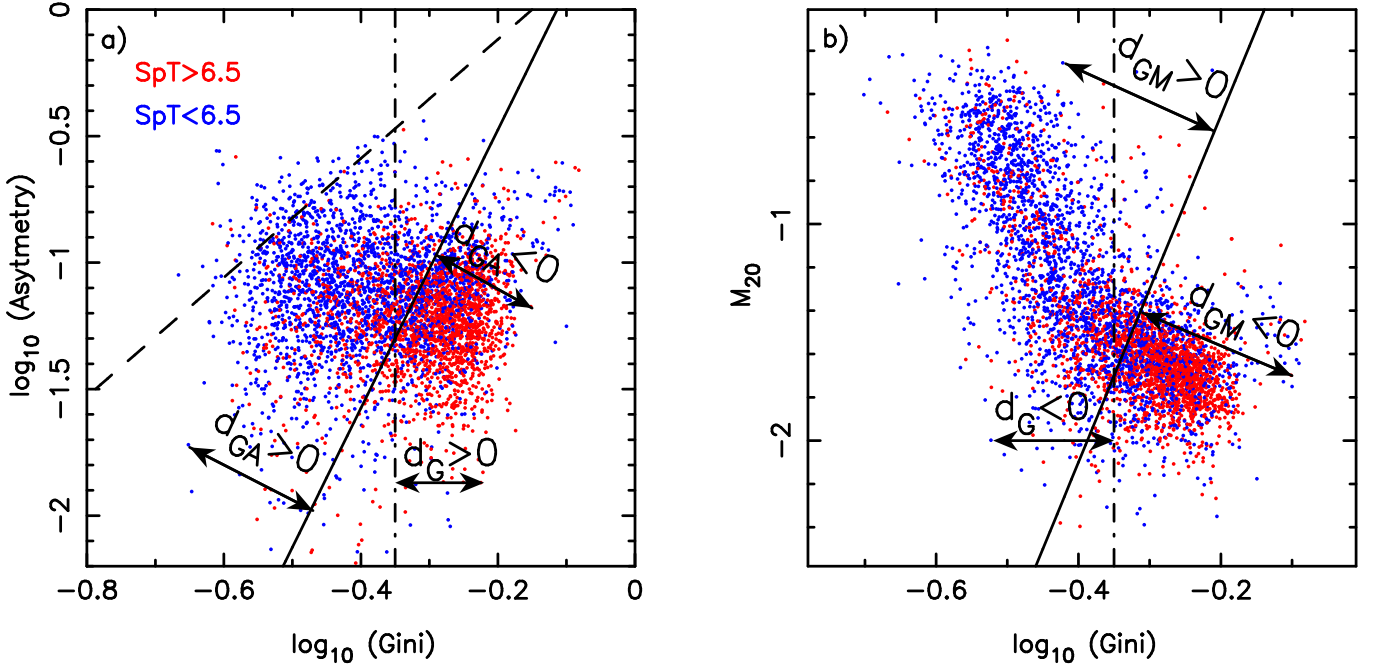}
\caption{
(a) Gini coefficient ($G$) vs. rotational asymmetry ($A$).  
Dividing lines are drawn between regions of
predominantly irregular, spiral, and elliptical types from
Capak et al. (2007b). The dashed line between irregular and
spiral galaxies is defined as $\log A = 2.353 \log G+0.353$,
while the solid line between spiral and early-type galaxies
is defined as $\log A=5.5\log G+0.825$. The dot-dashed line 
is a cut at $\log G=-0.35$, used to select early-type 
galaxies in this paper. 
(b) Gini coefficient ($G$) vs. \mt.
The solid line is defined as $\mt=8.0\log G+1.1$. EROs that 
have spectral type, $\SpT>6.5$, are colored red, and those with 
$\SpT<6.5$ are blue. $d$ ($\dGA$, $\dG$, $\dGM$) is 
the perpendicular distance from each point in the diagram 
to the dividing lines.
}
\label{fig:mor}
\end{figure*}

Capak et al. (2007b) have compared the morphology classification
for $\sim 2000$ F814W\, $< 22.5$ mag galaxies, by rotational
asymmetry and Gini coefficient, and by visual morphologies. They
found that both $\log G=-0.35$ and $\log A=5.5\log G+0.825$
can be used to classify early-type galaxies and late-type
galaxies using only the $HST$/ACS F814W images. 
Figure~\ref{fig:mor}(a) shows the distribution of EROs in the
$\log G$ versus $\log A$ (asymmetry index) plane. As in Capak et
al. (2007b), the dashed line is defined as
$\log A=2.353 \log G + 0.353$,
used to separate irregular and spiral galaxies;
the solid line is defined as $\log A=5.5 \log G+0.825$, used to
separate spiral and early-type galaxies; and the dot-dashed line
is defined as $\log G=-0.35$, used to separate early-type and
late-type galaxies. The classifications by the SED fitting
method are indicated with different colors in this figure,
OGs (with $\SpT>6.5$, early type) are shown as red, and DGs (with
$\SpT<6.5$, late type) are shown as blue. From this figure, we
found that the distribution of EROs in the $\log A$ versus $\log G$
diagram is in good agreement with classifications based on
visual method, most of the EROs with $\SpT>6.5$ have
$\log G>-0.35$ or $\log A<5.5\log G+0.825$.
If we chose a cut at $\log G=-0.35$ as the dividing line between
 early- and late-type galaxies, the fractions of OGs (early type) 
and DGs (late type) are 48\% and 52\%, respectively.  
If we chose a cut at $\log A=5.5\log G+0.825$ as the dividing 
line between early- and late-type galaxies, the fractions of OGs 
(early type) and DGs (late type) are 47\% and 53\%, respectively.

In Figure~\ref{fig:mor}(b), we show the Gini coefficient and 
\mt\, of EROs. The distribution of EROs is very similar to 
that of local galaxies in Lotz et al. (2004), with OGs 
showing high $G$ and low \mt\, values, and DGs with lower
$G$ and higher \mt\, values. Most of the OGs have
$\mt\, <8.0\log G+1.1$.
If we use this line as the selection limits for early-
and late-type galaxies, we found 47\% of EROs are classified
as OGs and 53\% of them are DGs.

\subsection{Classification based on Principal Component
Analysis}
\label{sec:pca}

The agreement among these different methods is found to
be satisfactory. Among the EROs with $ \Kv < 19.2$ mag,
$\sim$52\% of the EROs classified as DGs, and $\sim$48\% of
them classified as OGs. However, when we consider some EROs
in our sample, they may be classified as OGs by one method,
but be classified as DGs by the other methods. 
To check the efficiency of these different methods, and 
classify all EROs in our sample definitely, we therefore 
performed a PCA using the measurements of $\SpT$, $[3.6]-[8.0]$, 
$G$, $A$, and \mt\, as basic variables.

PCA is mathematically defined as an orthogonal linear
transformation that transforms the data to a new coordinate
system such that the greatest variance by any projection of
the data comes to lie on the first coordinate (called the first
principal component, PC$_1$), the second greatest variance on
the second coordinate (PC$_2$), and so on (Kong \& Cheng 2001;
Scarlata et al. 2007). The principal components (PCs) are a linear 
combination of the original variables and define a new coordinate 
system obtained by rigid rotation of the original space.
All variables are standardized before performing the analysis 
by subtracting their median value and normalizing them with 
their standard deviation ($\sigma$).
The five variables considered in this paper are defined
as: $x_1=(\dS-\bar{d}_S)/\sigma_{\dS}$, $\dS=\SpT-6.5$
($\SpT=6.5$ was used to classify OGs and DGs in Section 4.1);
$x_2=(\dI-\bar{d}_I)/\sigma_{\dI}$, 
$\dI=([3.6]-[8.0])-(-0.95)$
([3.6]-[8.0]$=-0.95$ was used to classify OGs and DGs in
Section 4.2); $x_3=(\dG-\bar{d}_G)/\sigma_{\dG}$, $\dG=\log
G-(-0.35)$ ($\log G=-0.35$ was used as the dividing
line between early- and late-type galaxies in Section 4.3);
$x_4=(\dGA-\bar{d}_{GA})/\sigma_{\dGA}$, where $\dGA$ is
the perpendicular distance from each point in the $\log G$ versus
$\log A$ diagram to the line $\log A=5.5 \log G+0.825$,
which was used to separate OGs and DGs in Section 4.3;
$x_5=(\dGM-\bar{d}_{\rm GM})/\sigma_{\dGM}$, where 
$\dGM$
is the perpendicular distance from each point in the $\log
G$ versus  $M_{20}$ diagram to the line $M_{20}=8.0 \log G+1.1$,
which was used to separate OGs and DGs in Section 4.3.

There are 5264 EROs in our sample, and all of them have
$\SpT$ parameters, but only 4761 of them have [3.6]-[8.0]
values (with fluxes brighter than the $3\sigma$ flux limit
of 1 $\mu$Jy at both 3.6 and 8.0 $\mu$m band), and 4455 of
them have nonparametric quantification ($G$, $A$, and \mt)
of galaxy morphology. Therefore, only 4028 EROs in our sample
have all $\SpT$, $[3.6]-[8.0]$, $G$, $A$, and \mt\, 
measurements, we apply the PCA method on these EROs, and the 
results are presented in Table~\ref{tab:pca}.  
The first row gives the
eigenvalue (i.e., variance) of the data along the direction
of the corresponding PC. The second row shows the fraction
of the variance that is explained by each of the PCs, i.e.,
the fraction of the "power" that is contained in each PC;
the third row lists the cumulative fraction of the variance.
In the last five rows of the table, each column lists
the weights assigned to each input variable, in the linear
combination that gives the direction of the specific PC. Some
distinct characters can be found from this table. First, the
first two PCs account for $\sim$82\% of the total variance,
the other PCs can be ignored. Therefore, to classify EROs,
we will use the first two PCs only. Second, for PC$_1$
(PC$_1=0.34x_1-0.26x_2+0.54x_3-0.54x_4-0.49x_5$), the absolute
weights of $x_3$, $x_4$, and $x_5$ are large; on the other hand,
for PC$_2$ (PC$_2=-0.56x_1+0.71x_2+0.28x_3-0.23x_4-0.22x_5$),
the absolute weights of $x_1$ and $x_2$ are large. Therefore,
PC$_1$ is dominated by $\dG$, $\dGA$, and $\dGM$, and PC$_2$
is dominated by $\dS$ and $\dI$.

\begin{table}
\centering
\caption{Results of the PCA method,
including the eigenvalue, fraction of eigenvalue, and weights
of each input variable.}
\label{tab:pca}
\vspace{0.1cm}
\begin{tabular}{lrrrrr}
\tableline
\tableline
Variable  & PC$_1$&  PC$_2$&  PC$_3$&  PC$_4$&  PC$_5$\\
\tableline
Eigenvalue\tablenotemark{a}    &   2.97&    1.11&    0.55&    0.34&    0.03\\
Proportion\tablenotemark{b}    &  59.42&   22.14&   10.94&    6.81&    0.69\\
Cumulative\tablenotemark{c}    &  59.42&   81.56&   92.50&   99.31&  100.00\\
\\
$\dS  (x_1)$   &   0.34&   -0.56&    0.75&   -0.09&    0.03\\
$\dI  (x_2)$   &  -0.26&    0.71&    0.64&   -0.07&   -0.02\\
$\dG  (x_3)$   &   0.54&    0.28&   -0.10&   -0.33&    0.72\\
$\dGA (x_4)$&  -0.54&   -0.23&    0.10&    0.41&    0.69\\
$\dGM (x_5)$&  -0.49&   -0.22&   -0.05&   -0.84&    0.05\\
\tableline
\end{tabular}
\tablenotetext{a} {The eigenvalue of the data along 
the direction of the corresponding PC.}
\tablenotetext{b} {The fraction of the variance that is contained in 
each PC.}
\tablenotetext{c} {The cumulative fraction of the variance.}
\end{table}

\begin{figure*}
\centering
\includegraphics[angle=-90,width=\textwidth]{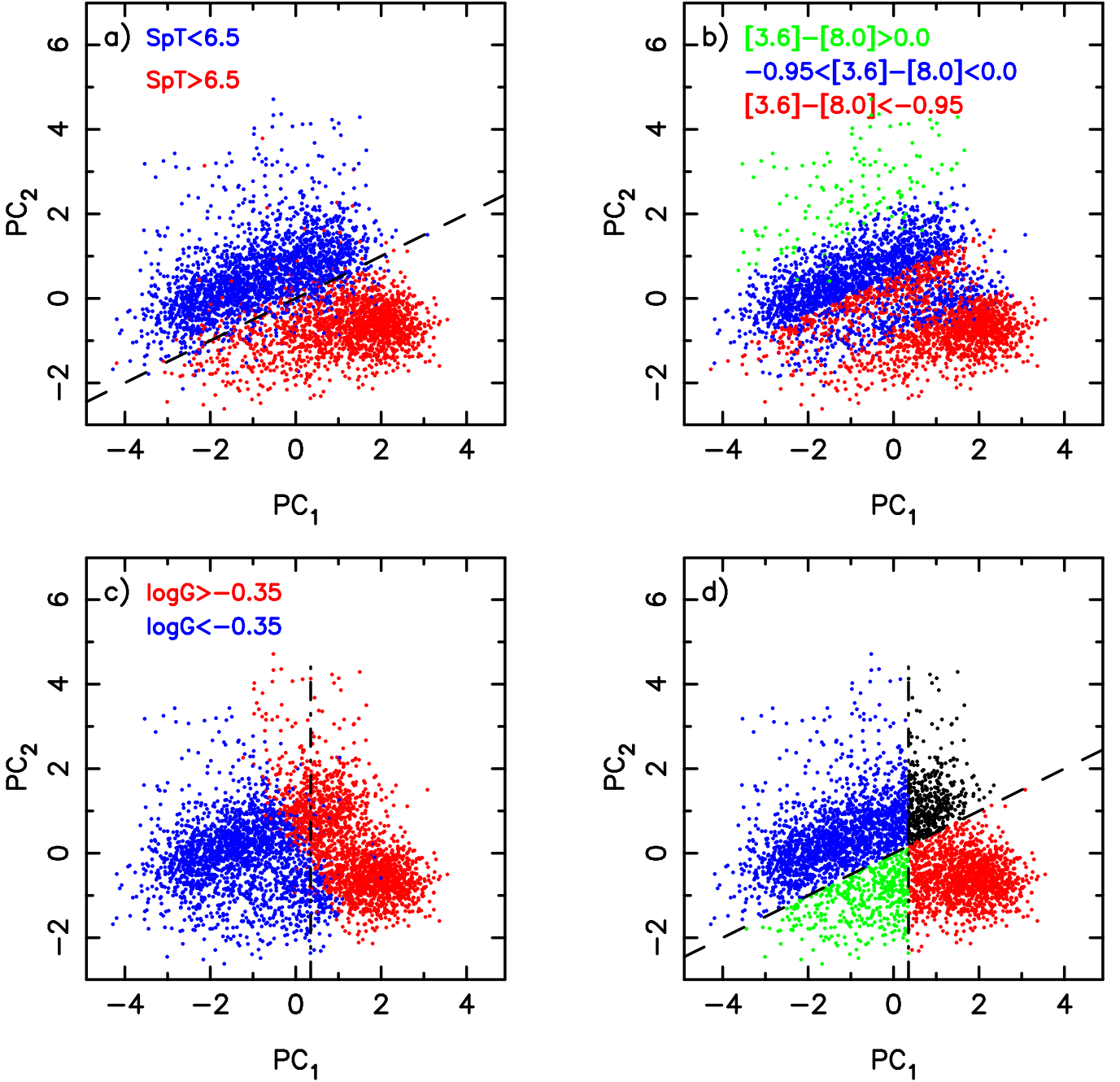}
\caption{
Plot of PC$_1$ vs. PC$_2$ for EROs in our sample.
(a) EROs are colored with $\SpT$, $\SpT>6.5$ (OGs) are red, 
DGs are blue.
(b) $[3.6]-[8.0]<-0.95$, $[3.6]-[8.0]>0.0$, and
$-0.95<[3.6]-[8.0]<0.0$ are colored as red, green, and 
blue, respectively.
(c) $\log G> -0.35$ (OGs) are plotted as red and DGs as 
blue.
(d) EROs are colored with the best-separated line from 
panel (a) and panel (c).
}
\label{fig:pca}
\end{figure*}

Figure~\ref{fig:pca} shows the EROs in our sample on PC$_1$
versus  PC$_2$ diagram, colored with different parameters.
In Figure~\ref{fig:pca}(a), EROs with $\SpT>6.5$ are shown as 
red color, but with $\SpT< 6.5$ are shown as blue color. 
In Figure~\ref{fig:pca}(b), EROs with $[3.6]-[8.0]<-0.95$
are shown as red, with $-0.95<[3.6]-[8.0]<0.00$ are shown
as blue, and with $[3.6]-[8.0]>0.00$ are shown as green.
Because PC$_2$ is dominated by $\dS$ ($x_1$) and $\dI$
($x_2$), and OGs have $\dS$>0 and $\dI<0$, so OGs have a 
small PC$_2$ value, but DGs have a large PC$_2$ value.  
From Figure~\ref{fig:pca}(a), we found that EROs with 
$\SpT>6.5$ and $\SpT<6.5$ can be separated by the dashed 
line PC$_2=0.5\times$PC$_1$. 
In Figure~\ref{fig:pca}(b), we found a similar sequence as in 
Figure~\ref{fig:pca}(a); AGNs stay at the upper-left area, OGs 
stay at the down-right area, and DGs stay at the middle area. 
The other character which can be found in Figure~\ref{fig:pca}(b) is 
that OGs and DGs are mixed in the separation area, which was 
caused by: [3.6]-[8.0] can be used to classify EROs only if those 
infrared photometry are measured accuracy; however, in 
practice the required photometric precision is often 
difficult to obtain. 
A clear classification of EROs, which lies within 0.1 
mag of the dividing line, is difficult.
In addition, the $[3.6]-[8.0]$ color also depends on reddening 
and redshift of galaxy.

\begin{figure*}
\centering
\includegraphics[angle=-90,width=\textwidth]{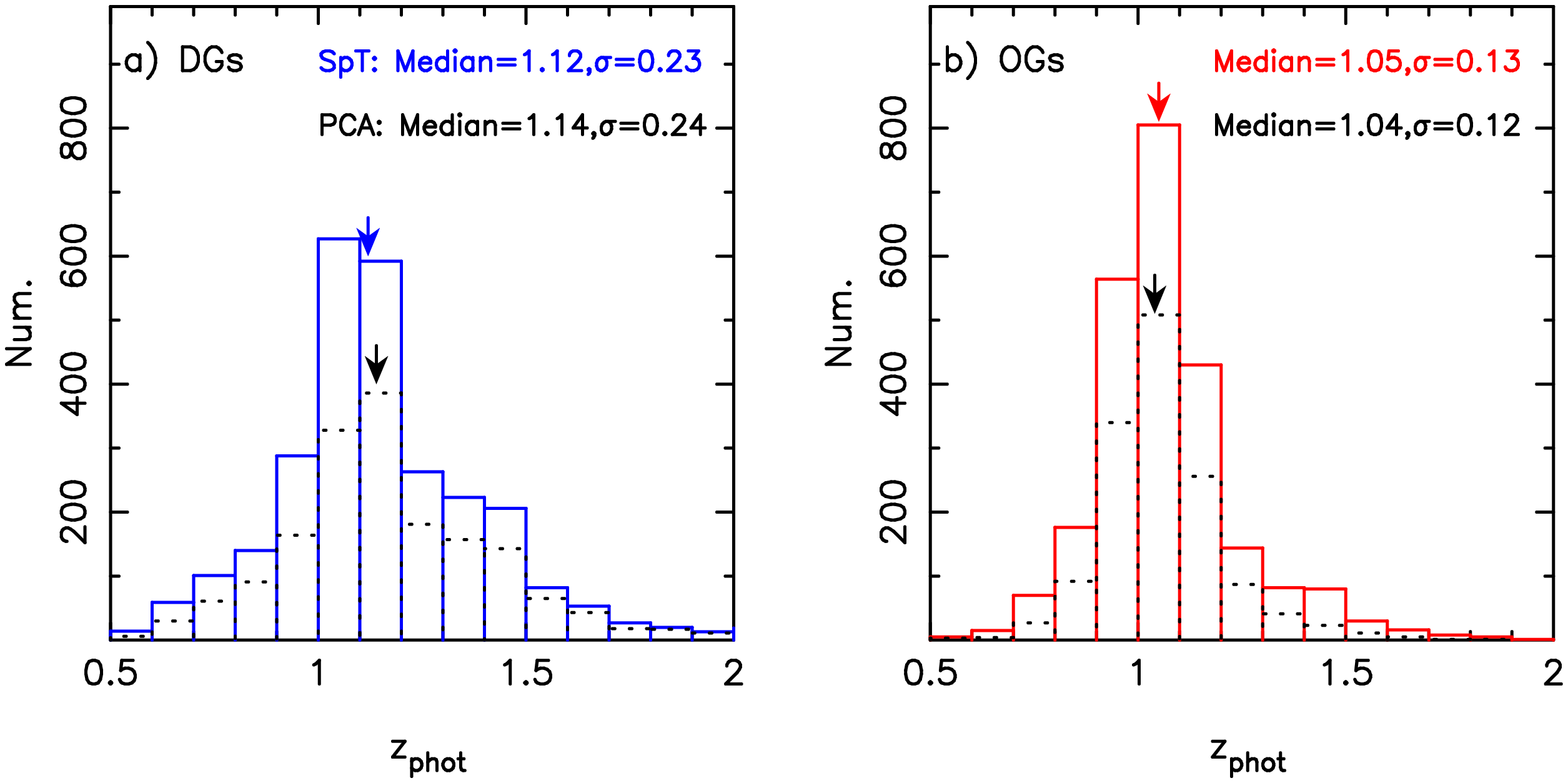}
\caption{
Redshift distribution of EROs in our sample. Panel (a) shows DGs.
Panel (b) shows OGs. The solid lines show the redshift distribution
of DGs and OGs, which were classified by the SED fitting method;
the dotted lines show the result by the PCA method.
}
\label{fig:distz}
\end{figure*}

In Figure~\ref{fig:pca}(c), EROs with $\log G>-0.35$ were 
plotted as red points, and with $\log G<-0.35$ as blue 
points. The distributions are similar if we plot EROs
with $\dGA<0$ or $\dGM<0$ as red, and with $\dGA>0$ or
$\dGM>0$ as blue. PC$_1$ is dominated by $G$, $A$, and \mt.
Comparing with DGs, OGs have larger $\dG$ (weight value is
0.54, positive, see Table~\ref{tab:pca} for detail), smaller
$\dGA$ (-0.54, negative) and $\dGM$ (-0.49, negative),
so the PC$_1$ of OGs are larger than those of DGs. 
EROs with $\log G >-0.35$ and $\log G <-0.35$ can be 
separated by the dot-dashed line as PC$_1=0.30$. 
Using the dashed line in Figure~\ref{fig:pca}(a) and the 
dot-dashed line in Figure~\ref{fig:pca}(c), we divide PC$_1$ 
versus PC$_2$ diagram into four different district, and plot 
EROs with different colors in Figure~\ref{fig:pca}(d). 
Forty-three percent of EROs with PC$_2>0.5$PC$_1$ and PC$_1<0.30$ are 
classified as DGs (blue points), they have late-type SEDs and 
small concentrations.  
Thirty-six percent of EROs in our sample with PC$_2<0.5$PC$_1$ and 
PC$_1>0.30$ are classified as OGs (red points), they have 
early type SEDs and high concentrations.  
Eleven percent of them have PC$_2>0.5$PC$_1$ and PC$_1>0.30$ (black 
points), so they have late-type SEDs, but elliptical type 
morphologies. This kind of EROs has been reported by 
Cotter et al. (2005). Using near-infrared spectroscopy, 
ground-based optical and near-infrared imaging, and $HST$ imaging, 
Cotter et al. found vigorous star formation in a bulge-dominated 
ERO at $z= 1.34$. The other EROs (10\%) have PC$_2<0.5$PC$_1$, but PC$_1<0.30$
(green points).  These EROs have late-type morphologies, but
have early-type SEDs ($\SpT>6.5$, but $\SpT$ for most of them
are less than 8, well fitted with intermediate templates),
probably many of them are late-type galaxies, but have
large underlying bulges, and so their SEDs may naturally be
dominated by underlying old stellar populations. Stockton et
al.(2006) have found those kinds of EROs in two quasar fields
at $z\sim1.4$; SEDs and rest-frame
near-UV spectra of them show that they are strongly dominated
by old stellar populations; radial surface brightness profiles
from adaptive optics images of these galaxies are best fitted by
profiles close to exponentials.

\section{Results and Discussion}

In this section, we briefly discuss the properties of OGs
and DGs found in our data, including redshift distributions,
number counts, and clustering of these two types EROs.

\subsection{Redshift distributions}\label{sec:distz}

Figure~\ref{fig:distz} shows the redshift distribution of DGs 
(in panel (a)) and OGs (in panel (b)) in our sample, as the solid 
lines, which were classified by the SED fitting method. 
The redshift distribution of DGs in our sample has a median 
value $z\sim1.12$ (this value depends on the magnitude limit 
of the observational survey, here for \Kv=19.2) and 
$\sigma=0.23$, with a tail toward higher redshift. 
The redshift distribution of OGs in our sample has a median 
value $z\sim1.05$ and $\sigma=0.13$. The peak of the redshift 
distribution of OGs in our sample is less than that of the DGs. It 
is worth noting that our OGs have a narrow redshift 
distribution, the sigma value of which is $\sigma=0.13$. Unlike 
the distribution of DGs, only few OGs have redshift higher 
than 1.4, the majority of OGs are located at $0.8<z<1.4$.
The dotted lines in Figure~\ref{fig:distz} show the redshift
distribution of our OGs and DGs, which were classified by the
PCA method. We found that the redshift distributions from these two
kinds of classification methods are similar; OGs in our sample
have small median value and a small sigma value.

\subsection{Number counts}\label{sec:dnum}

\begin{figure*}
\centering
\includegraphics[angle=-90,width=\textwidth]{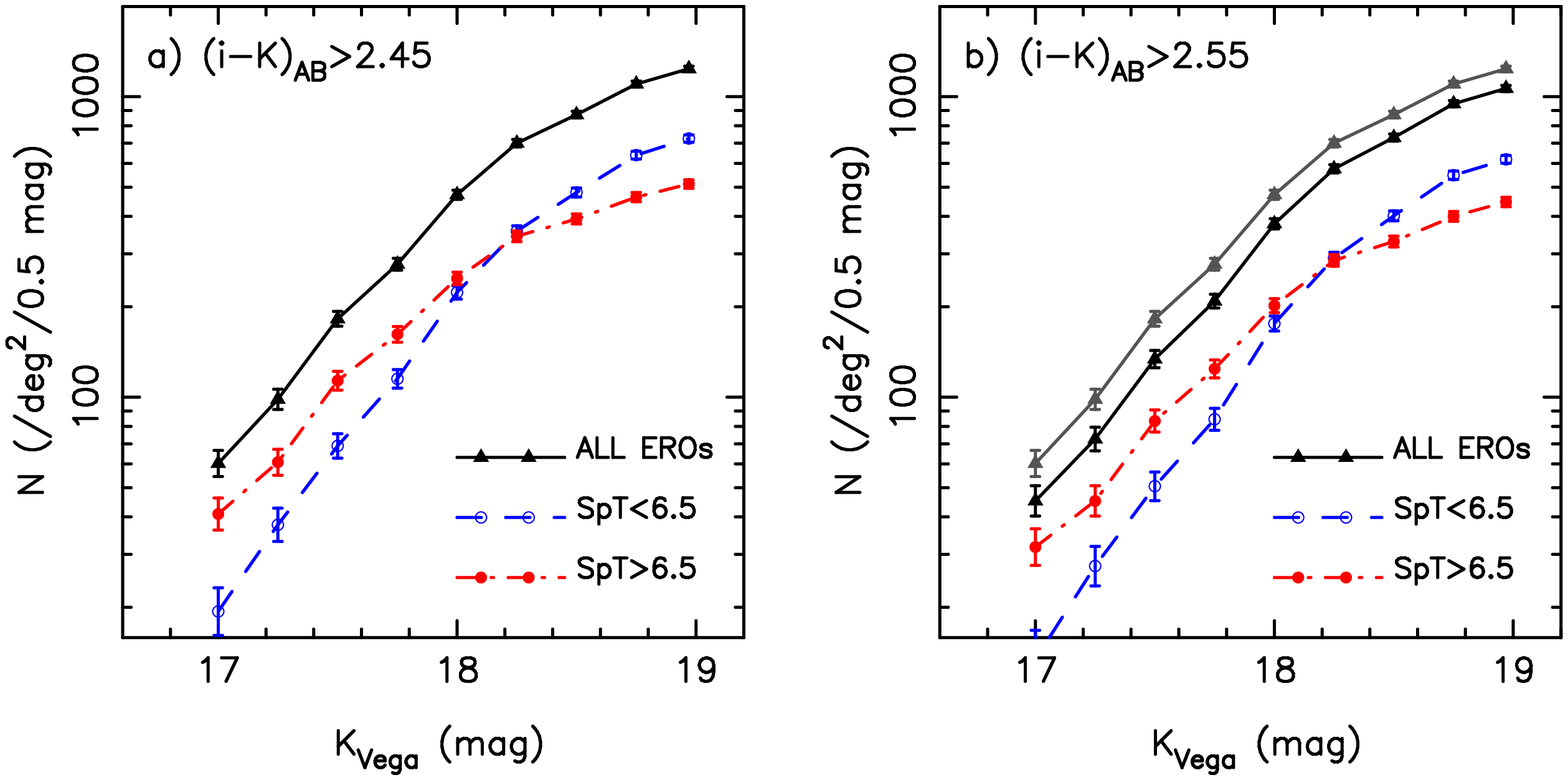}
\caption{
K-band differential galaxy number counts for OGs and DGs,
compared with the entire EROs sample as in 
Figure~\ref{fig:numc}.
The solid curves with triangles show the number counts for
EROs in the COSMOS, and the filled circles and open circles show the
number counts for OGs and DGs, respectively.
(a) EROs defined by color cut $(i-K)_{\rm AB} \geq 2.45$;
(b) EROs defined by color cut $(i-K)_{\rm AB} \geq 2.55$.
The error bars indicate the Poissonian uncertainties.
}
\label{fig:enum}
\end{figure*}

Number-magnitude relations, commonly called number counts,
provide a statistical probe of both the space distribution
of galaxies and its evolution.  
For this reason, we derived $K$-band differential number 
counts for DGs and OGs, classified by the SED fitting 
method, in the COSMOS field, and plotted them in 
Figure~\ref{fig:enum}(a). Also shown are the number counts
for all EROs in our sample, same as the symbols (triangles
with solid line) in Figure~\ref{fig:numc} for comparison.
To check how the color criterion influences the number 
counts of DGs and OGs, Figure~\ref{fig:enum}(b) shows the
number counts of EROs defined by color cut 
$(i-K)_{\rm AB} \geq 2.55$. The gray solid line with 
triangles in Figure~\ref{fig:enum}(b) represents EROs 
defined by color cut $(i-K)_{\rm AB} \geq 2.45$.

The open circles with the dashed line in Figure~\ref{fig:enum} show
the number counts for DGs ($\SpT<6.5$), the filled circles with the 
dot-dashed line show the number counts of OGs ($\SpT>6.5$)
in the COSMOS. We find that the relative fractions of EROs strongly 
depend on the K-band limited magnitudes, and the fraction of DGs
in the COSMOS field increases very steeply toward fainter
magnitudes: 35\% at \Kv=17.5 mag, 39\% at \Kv=18.0, 46\%
at \Kv=18.5, and 52\% at \Kv=19.2. 
For OGs, the slope of the number counts is variable, being 
steeper at bright magnitudes and flattening out toward faint 
magnitudes. A break in the counts is present at $\Kv\sim 18.5$, 
very similar to the break in the number counts for the entire 
EROs sample. The counts of DGs have roughly the same slope at 
all K-band magnitudes.
These results are due to the narrow redshift range of OGs, but
the much wider redshift distribution of DGs, as shown in 
Figure~\ref{fig:distz}. 
From Fig.~\ref{fig:enum}(a) and Figures~\ref{fig:enum}(b), we 
find that the color criterion has little or no influence on the 
number counts of DGs and OGs.

\subsection{Color versus photometric redshift}

\begin{figure}
\centering
\includegraphics[angle=-90,width=\columnwidth]{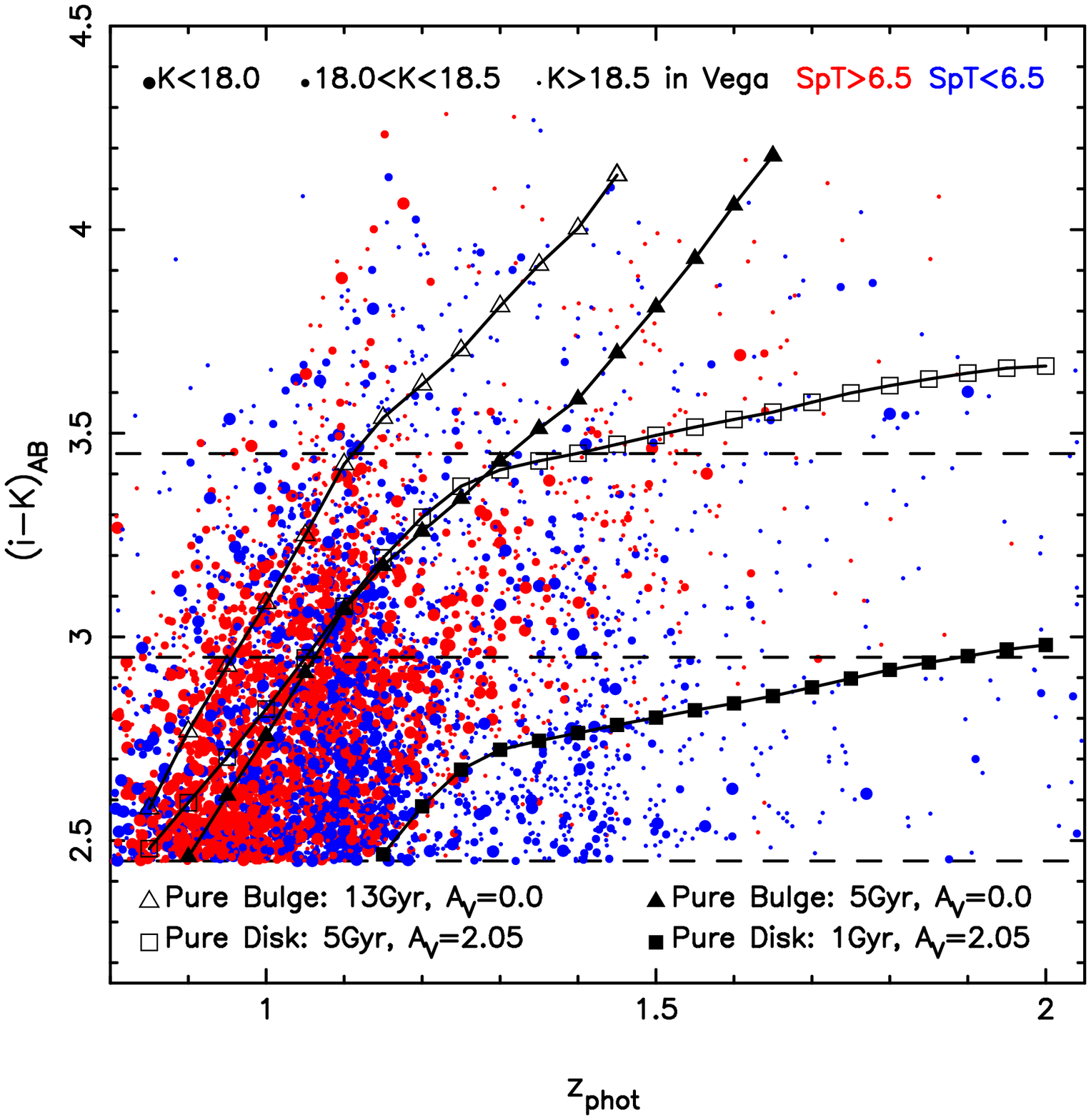}
\caption{
Color vs. photometric redshift distribution for EROs in the
COSMOS.  Red circles correspond to OGs, and blue circles 
correspond to DGs.  
Symbol size is keyed to the K-band magnitude. The four solid 
lines indicated predications by passive evolution and dusty 
starburst models.
}
\label{fig:zik}
\end{figure}

Figure~\ref{fig:zik} shows the  $(i-K)_{\rm AB}$ color as a
function of photometric redshift for the EROs in the COSMOS
field. The red and blue points correspond to OGs and DGs,
which are classified by the SED fitting method.

Variety of color criteria were used to define EROs, for example
$R-K \geq 5$, 5.3, or 6.  To check how the color criterion
influences the fraction of EROs, we use  $(i-K)_{\rm AB} \geq
2.45$, 2.95, and 3.45 for EROs' selection (dashed horizontal
lines in Figure~\ref{fig:zik}).  As described in Section 3, we
used $(i-K)_{\rm AB} \geq 2.45$ to select the entire EROs sample
in this paper, and found $\sim52$\% of them were classified as
DGs, and $\sim48$\% of them as OGs.  If we use $(i-K)_{\rm AB}
\geq 2.95$ as the color criterion, the fraction of DGs is 53\%.
However, if we use $(i-K)_{\rm AB} \geq 3.45$, we found the
fraction of DGs increases to 61\%.

Figure~\ref{fig:zik} shows that the distribution of OGs/DGs is
very inhomogeneous in the color versus photometric redshift
diagram. As shown in Section~\ref{sec:distz}, DGs have a higher
redshift tail, and OGs have a narrow redshift distribution. For
EROs with $\zph>1.2$, $\sim$ 73\% of them are DGs.  To compute
the expected $(i-K)_{\rm AB}$ versus redshift of different kinds
of dust-free passively evolving galaxies (pure bulge) and dusty
starburst galaxies (pure disk), we used the KA97 model.  The filled-
and open-triangles represent passively evolving galaxies with
ages of 5 Gyr and 13 Gyr, and the filled- and open-squares represent
dusty ($A_{\rm V}=2.05$) starburst galaxies with ages of 1 Gyr
and 5 Gyr, respectively.  From the tracks of stellar population
synthesis model, we found that the dust-free passively evolving
galaxies have a narrow redshift range, EROs with $z>1.2$ and
$(i-K)_{\rm AB}<3.2$ are found to be located outside the region
of pure bulge galaxies. On the other hand, the expected color
for starburst galaxies has a wide range of redshifts.

\subsection{Clustering of the EROs}

Measuring the clustering of galaxies provides an additional
tool for studying the evolution of galaxies and the formation
of structures (Kong et al. 2006; Heinis et al. 2007; 
McCracken et al. 2007).  We estimate the clustering
properties of the entire EROs, OGs, and DGs (classified by 
the SED
fitting method), using the estimator defined by Landy \&
Szalay (1993).  
In our analysis a fixed slope $\delta=0.8$ was assumed for 
the two-point correlation function,
 $w(\theta)=A\times\theta^{-\delta}$. 

\begin{figure*}
\centering
\includegraphics[angle=-90,width=0.95\textwidth]{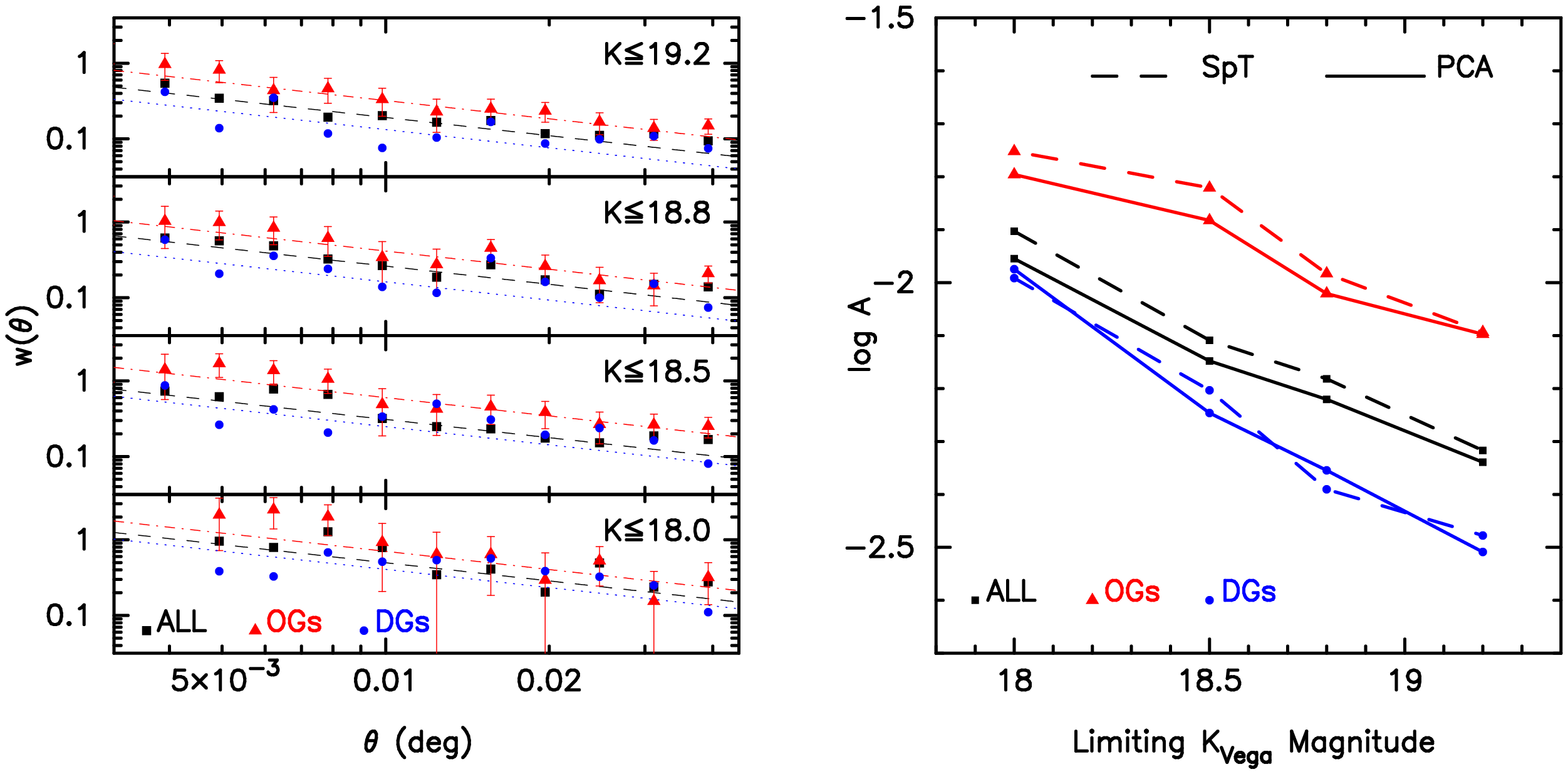}
\caption{
Left panel: the observed, bias-corrected two-point correlations
of the entire EROs (squares), OGs (triangles), and DGs
(circles). The error bars on the direct estimator values
are $1 \sigma$ errors. To make this plot clear, we show
the error bars of OGs only. Because of the small number of
objects included, some bins were not populated. 
Right panel: the angular clustering amplitudes of the entire 
EROs, OGs, and DGs are shown as a function of the $K$-band 
limiting magnitudes of the sample analyzed. 
The solid and dashed lines represent the sample were classified 
by the PCA method and the SED fitting method.
}
\label{fig:clustering}
\end{figure*}

In the left panel of Figure~\ref{fig:clustering}, the 
bias-corrected two-point correlation functions $w(\theta)$ 
of the entire EROs, OGs, and DGs are shown as squares, 
triangles, and circles, respectively.
The dashed line shows the power-law correlation function
given by a least-square fit to the measured correlations. 
The derived clustering amplitudes (where $A$ is the amplitude
of the true angular correlation at $1^\circ$) are presented in 
Columns 2-4 of Table~\ref{tab:clustering}, and shown in 
the right panel of Figure~\ref{fig:clustering} as dashed lines.
Furthermore, by using the Limber's equation (Limber 1954) and 
the photometric redshift distributions of the entire EROs, OGs, 
and DGs, we also derived their comoving correlation lengths 
from the angular clustering amplitudes, at the same K-band
 limits. 
The spatial correlation lengths $r_0$ for the entire EROs, OGs, 
and DGs are listed in Columns 5-7 of 
Table~\ref{tab:clustering}.

\begin{table*}
\centering
\tiny
\caption{Clustering Amplitudes $A$ (10$^{-3}$) of $w(\theta)$ at 1$^\circ$ 
and spatial correlation lengths $r_{\rm 0}$ in $h^{-1}$ Mpc 
for the entire EROs, OGs, and DGs Samples in the COSMOS}
\label{tab:clustering}
\vspace{0.2cm}
\begin{tabular*}{1.0\textwidth}{@{\extracolsep{\fill}}lrrrrrrrrrrrrr}
\tableline
\tableline
       &\multicolumn{6}{c}{Classified by the SED Fitting
       Method}& &\multicolumn{6}{c}{Classified by the PCA
       method}\\
\cline{2-7} \cline{9-14}

$K$ Limit & EROs ($A$)&OGs($A$)& DGs($A$)&EROs($r_0$)& OGs($r_0$) & DGs($r_0$)       
&& EROs($A$)& OGs($A$)& DGs($A$)&EROs($r_0$)& OGs($r_0$) & DGs($r_0$)\\
\tableline
18.0 & 12.5$\pm$0.7&17.7$\pm$0.8&10.2$\pm$4.4& 13.6$\pm$0.4&14.7$\pm$0.3&13.0$\pm$3.1&&
       11.1$\pm$1.1&16.0$\pm$1.6&10.6$\pm$8.2& 12.7$\pm$0.7&14.4$\pm$0.7&12.6$\pm$5.5\\
18.5 &  7.8$\pm$0.2&15.1$\pm$0.2& 6.3$\pm$1.3& 10.5$\pm$0.2&12.6$\pm$0.1&10.7$\pm$1.1&&
        7.1$\pm$0.3&13.1$\pm$0.5& 5.7$\pm$2.7& 10.0$\pm$0.3&11.1$\pm$0.2&10.4$\pm$2.5\\
18.8 &  6.6$\pm$0.1&10.4$\pm$0.2& 4.1$\pm$0.9&  9.6$\pm$0.1&10.2$\pm$0.1& 8.4$\pm$1.0&&
        6.0$\pm$0.2& 9.5$\pm$0.5& 4.4$\pm$1.9&  9.1$\pm$0.2& 9.3$\pm$0.3& 9.0$\pm$1.9\\
19.2 &  4.8$\pm$0.1& 8.1$\pm$0.1& 3.3$\pm$0.5&  8.0$\pm$0.1& 8.9$\pm$0.1& 7.5$\pm$0.6&&
        4.6$\pm$0.2& 8.0$\pm$0.4& 3.1$\pm$1.3&  7.8$\pm$0.2& 8.4$\pm$0.2& 7.4$\pm$1.5\\
\tableline
\end{tabular*}
\end{table*}

We find significant clustering for limiting $K_{\rm Vega}$ 
magnitudes between 18.5 and 19.2 (Figure~\ref{fig:clustering} 
and Table~\ref{tab:clustering}). The clustering amplitudes
for our ERO sample are consistent with other previous 
studies (such as Firth et al. 2002; Georgakakis et al. 2005; 
Kong et al. 2006), and are much greater than the amplitudes
which have been measured for the field population at the 
same $K_{\rm Vega}$ limits (e.g., Kong et al 2006).  
The amplitudes shown in Figure~\ref{fig:clustering} suggest a 
trend of decreasing strength of the clustering for fainter 
EROs (was also found by Brown et al. 2005), OGs, and DGs.  
In addition, there appears to be a difference between the 
clustering properties of the OGs and DGs; the amplitudes and 
correlation lengths of the clustering of DGs are found to 
be much lower than those of OGs, a significant overdensity, 
by a factor of $\sim2$, of those OGs clustering amplitudes 
compare to DGs in the COSMOS field was found.

In the right panel of Figure~\ref{fig:clustering}, the solid
lines show the angular clustering amplitudes of the entire 
EROs, OGs, and DGs, which were classified by the PCA method.  
Since the cover area of $HST$ image is smaller than that of 
ground-based images, only 4028/5264 EROs in our sample have 
all $\SpT$, $[3.6]-[8.0]$, $G$, $A$, and \mt\, measurements, 
so the entire ERO sample here is different from that in 
the previous.
We calculated the clustering properties of the 4028 EROs,
DGs, and OGs, in the same manner as for the 5264 EROs. The
clustering properties of EROs, OGs, and DGs, by these two
classified method, are very similar.

\section{Summary}

In order to study the properties of galaxies at redshift $
z \sim 1$, we constructed a large sample of EROs 
from the multi-wavelength data (from CFHT-$u^\star$ to
IRAC-8.0$\mu$m, with 13 bands) of the COSMOS field. We detected
5264 EROs with $i-K \ge 2.45$ down to $\Kv= 19.2$ in the data.

By fitting template spectra of evolutionary population synthesis
model to the multi-wavelength data, we classified the EROs
by their spectral types and estimated their redshifts. Our
photometric redshifts fit the spectroscopic redshifts for a
sample of 38 EROs from zCOSMOS well, with an average
$\delta z/(1+z_{spec})= 0.02$.
The redshift distribution of EROs has
a peak at $ z\sim 1.1$, with a range from $\sim$0.8 to 1.5.

By our template fitting method, we found that the relative
fraction of DGs is $\sim 52$\% ($\SpT<6.5$), $\sim48\%$ of the
EROs in our sample belong to the OG class ($\SpT>6.5$). Among
the OGs, 28\% was fitted by templates with $6.5 < \SpT \le10$,
it suggests that a significant fraction of OGs have a
non-negligible amount of star formation.

Using IRAC color $[3.6]-8.0]=-0.95/0.0$, we classified EROs as
OGs, DGs, and AGNs, the relative fraction of them is 45\%, 51\%,
and 4\%, respectively. These fractions agree with the results
of the template fitting method, in general. In addition, we
found that lots of EROs lie within 0.1 mag of the dividing
line, so a clear classification of EROs as OGs and DGs by the
$[3.6]-8.0]$ color requires very high photometric precision.

Using the $HST$/ACS F814W images, we have measured the
concentration ($C$), asymmetry ($A$), as well as the Gini
coefficient ($G$) and the second-order moment of the brightest 20\%
of their light (\mt) of EROs.  As found for nearby galaxies,
EROs follow a tight $G-\mt-C-A$ sequence, OGs have high Gini
coefficients and concentrations and low second-order moments
as a result of their bright and compact bulges, DGs have lower
Gini coefficients and concentrations and higher second-order
moments. Using the same criteria as nearby galaxy, we classified
EROs as OGs and DGs by their morphological parameters; the
fractions of OGs and DGs are similar to those of the previous
two methods.

To classify EROs in our sample definitely, and reduce
the redundancy of these different classification methods,
we performed a PCA on the measurements of $\SpT$, $[3.6]-[8.0]$,
 $G$, $A$, and \mt. We
found that the first two PCs full describe
the key aspects of the EROs classification.  Using the first PC
as the classification criterion, 48.3\% EROs in our sample is
classified as OGs, and 51.7\% as DGs.  If the first two PCs were
used as the classification criteria, EROs can be separated as:
36\% of them are OGs (early-type SED and early-type morphology);
10\% of them have late-type morphologies, but early-type SEDs;
11\% of them have late-type SEDs, but elliptical type
morphologies; 43\% of them are DGs (late-type SED and late-type morphology).

Finally, we examined the properties of OGs and DGs in our
sample, and found that: DGs have a wide range of redshift,
with a peak at $z\sim1.12$ and $\sigma=0.23$, the redshift
distribution of OGs in our sample has a median value 
$z\sim1.05$
and $\sigma=0.13$; the relative fractions of EROs strongly 
depend on the K-band limited magnitudes, the fraction of
DGs increases very steeply toward fainter magnitudes; the
angular clustering amplitude of OGs is a factor of $\sim2$
larger than that of DGs. Moreover, the clustering strength 
of EROs, OGs, and DGs increases with the K-band limited 
magnitude decreases.

\acknowledgments
We thank an anonymous referee for useful and constructive 
comments that resulted in a significant improvement of this 
paper.
It is a pleasure to thank S. White and E. Daddi for useful
suggestions. We gratefully acknowledge R. Abraham for 
access to his morphology analysis code and P. Capak for his 
help on COSMOS data. The $HST$ COSMOS Treasury program 
was supported through NASA grant HST-GO-09822.  
The work is supported by the National Natural Science 
Foundation of China (NSFC, Nos. 10573014, 10633020, and 10873012), 
the Knowledge Innovation Program of the Chinese Academy of 
Sciences (No. KJCX2-YW-T05), and National Basic Research 
Program of China (973 Program; No. 2007CB815404).


\end{document}